\documentclass[preprint,showpacs,aip,pop,amsmath]{revtex4-1}
\usepackage{graphicx}
\usepackage{amsmath}
\usepackage{amssymb}
\usepackage{color}
\usepackage{url}
\usepackage{subfig}
\usepackage{bm}
\usepackage{verbatim }
\usepackage{pifont}
\renewcommand{\vec}[1]{\boldsymbol{#1}}

\newcommand{\eps}{\varepsilon}
\renewcommand{\d}{\mathrm{d}}

\renewcommand{\vec}[1]{{\mathbf{#1}}}
\newcommand{\dx}{\,\mathrm{d}x}

\newcommand{\Eins}{\mathbf{1}}

\newcommand{\bhat}{\bm{\hat{b}}}

\newcommand{\esc}[3]{${#1}$E{#2}${#3}$}
\begin{document}

\preprint{}

\title{Three discontinuous Galerkin schemes for the anisotropic heat conduction equation on
non-aligned grids}
\author{M.~Held}
\email{Markus.Held@uibk.ac.at}
\author{M.~Wiesenberger}
\affiliation{Institute for Ion Physics and Applied Physics, University of
   Innsbruck, Technikerstrasse 25, A-6020 Innsbruck, Austria}
\author{A.~Stegmeir}
\affiliation{Max--Planck--Institut f\"ur Plasmaphysik, Boltzmannstrasse 2, 85748 Garching, Germany}
 
\begin{abstract}
We present and discuss three discontinuous Galerkin (dG) discretizations for the anisotropic heat conduction equation on non-aligned cylindrical grids. Our non-aligned scheme relies on a 
self-adjoint local dG (LDG) discretization of the elliptic operator. It conserves the energy exactly and converges with  arbitrary order. 
The pollution by numerical perpendicular heat fluxes decreases with superconvergence rates. 
We compare this scheme with aligned schemes that are based on the flux-coordinate independent approach for the discretization of parallel derivatives. 
Here, the dG method provides the necessary interpolation. The first aligned discretization can be used in an explicit time-integrator. 
However, the scheme violates conservation of energy and shows up stagnating convergence rates for very high resolutions. 
We overcome this partly by using the adjoint of the parallel derivative operator to construct a second self-adjoint aligned scheme. 
This scheme preserves energy, but reveals unphysical oscillations in the numerical tests, which result in a decreased order of convergence. 
Both aligned schemes exhibit low numerical heat fluxes into the perpendicular direction and are superior for flute-modes with finite parallel gradients. 
We build our argumentation on various numerical experiments on all three schemes for a general axisymmetric magnetic field, 
which is closed by a comparison to the aligned finite difference (FD) schemes of References~\cite{Stegmeir2014,Stegmeir2015}.
\end{abstract}
\maketitle


\section{Introduction}\label{sec:introduction}
Strong magnetic fields are known to separate the scales of turbulent motion in thermonuclear fusion or astrophysical plasmas.
Along and perpendicular to magnetic field lines the mean free paths are of the order of the connection length and of
the gyroradius. This introduces a strong anisotropy into the magnetized plasma, resulting in viscosity and heat conduction coefficients,
which are several magnitudes higher in the parallel direction than into the perpendicular direction. \\
In order to formulate the problem we split the gradient into parallel and perpendicular parts via 
\begin{align}
    \vec{\nabla}  =: \bhat \nabla_\parallel + \vec{\nabla}_\perp,
    \label{}
\end{align}
where $\nabla_\parallel = \bhat\cdot \nabla$, and $\bhat(\vec x) = \vec{B}/B$ is the unit vector in the magnetic field direction.
With this definition we can split the diffusion operator into
\begin{align}
    \Delta f = \nabla^2 f= \nabla\cdot(\bhat \bhat\cdot\nabla + \nabla_\perp)f  =: \Delta_\parallel f + \Delta_\perp f.
    \label{eq:definition}
\end{align}
In the following we mean to study the exemplary model of the anisotropic heat conduction equation 
\begin{align}
    \frac{\partial}{\partial t} T = \chi_\parallel \Delta_\parallel T +  \chi_\perp \Delta_\perp T, 
    \label{eq:diffusion}
\end{align}
with constant parallel and perpendicular heat conduction coefficients $\chi_\parallel$ and $\chi_\perp$. 
Since in magnetized plasmas the parallel heat conduction is orders of magnitudes higher than the perpendicular heat conduction, we assume that $\chi_\perp$ is zero. 
The energy \(E = \int_\Omega dV \hspace{1 mm} T  \) is invariant over time 
if the surface integral \(\int_{\partial \Omega} d\vec{A} \cdot  \vec{b} \nabla_\parallel T   = 0\) on the boundary
\(\partial \Omega\) of the domain \(\Omega\).
\\
The anisotropy imposes stringent requirements on the numerical method, which is commonly solved by aligning the coordinates with the magnetic field~\cite{Dhaeseleer1990}.
However, aligned flux coordinates exhibit nonorthogonal and anisotropic meshes while not resolving singular points (e.g.: X- and O-point).
Especially near the last closed flux surface of fusion plasmas those points result into severe numerical problems, which could only partly overcome by additional coordinate transformations~\cite{Scott2001,Ribeiro2010}. 
Here, the shifted metric transformation~\cite{Scott2001} remedies the nonorthogonality and the conformal 
tokamak coordinates~\cite{Ribeiro2010} produce isotropic meshes for a proper treatment of shaped nonsingular flux surfaces.
\\
The contrary approach is to misalign the numerical grid to the magnetic field.
Consequently one has to tackle numerical issues, 
which are most significantly convergence loss ~\cite{Babuska1992} and numerical perpendicular heat flux that exceeds the true perpendicular heat transport~\cite{Guenter2005}.
Hence, proper and more sophisticated numerical discretization techniques are unavoidable.
In general we can distinguish between non-aligned and aligned discretizations. 
The former methods have proven successful for finite differences~\cite{Guenter2005} and higher-order finite elements~\cite{Sovinec2004,Guenter2007}
even for huge levels of anisotropy $\chi_\parallel / \chi_\perp = 10^9$. The symmetric difference scheme of \cite{Guenter2005} is founded on the support operator method (SOM) introduced in Reference~\cite{Shashkov1995}, which is also
known as Mimetic Finite Difference (MFD) method. Its main idea is to define the divergence operator as the adjoint of the gradient operator, resulting in a self-adjoint diffusion operator.
This concept has also reached LDG methods, which are effectively used to solve parabolic and elliptic equations~\cite{Cockburn1998}. However, so far the properties of LDG method have not been investigated for 
anisotropic heat conduction in magnetized fusion plasmas.
\\
Aligned schemes rely either on 2D~\cite{Stegmeir2014,Stegmeir2015,Hariri2013} or 3D interpolation~\cite{Van_es_14}, and fix either the toroidal angle \(\Delta \varphi\) or the length of the field line \(\Delta s\). 
The interpolation points are found by tracing the magnetic field line with an appropriate integration method. Here, less accurate schemes neglect the detailed curvature of the magnetic field line and
employ an approximate aligned stencil. \\
The flux-coordinate independent (FCI) approach is an aligned 2D interpolation scheme, which approximates the parallel gradient \(\nabla_\parallel\) 
with a central finite difference~\cite{Hariri2013}.
In that case, high-order 2D interpolation techniques have effectively reduced numerical perpendicular diffusion. Here, a parallel advection equation supported the usability of the FCI approach even 
for diverted axisymmetric magnetic fields~\cite{Hariri2014}.
The FCI approach was adapted to the MFD method to tackle the anisotropic diffusion equation~\cite{Stegmeir2014}. 
It was shown that the self-adjoint discretization of the diffusion operator exhibits less numerical perpendicular diffusion than a naive discretization~\cite{Stegmeir2014,Stegmeir2015,stegmeir:phd15}. 
In this paper we lay the focus on the numerical investigation of the convergence rate of 
the aligned finite difference schemes for magnetic fields with \(\vec{\nabla} \cdot \bhat \neq 0\), which has not yet been considered.
\\
Aligned differences with 3D interpolation can be extended to yield comparable
numerical perpendicular heat fluxes to non-aligned finite differences in the case of magnetic fields with \(\nabla \cdot \bhat = 0\)~\cite{Van_es_14}. 
However, for \(\vec{\nabla} \cdot \bhat \neq 0\) and closed field lines along the symmetry axis it was unfolded that convergence is lost for high levels of anisotropy.
\\
The emphasis of all the so far mentioned schemes is put on magnetized fusion plasmas where only low temperature gradients along the magnetic field are expected. An extensive discussion on monotonicity preserving schemes,
which are eligible for astrophysical plasmas can be found in Reference~\cite{Sharma2007}.
\\
In this work we present a non-aligned scheme based on the local discontinuous Galerkin (LDG) method. 
This scheme is formulated in the spirit of the SOM~\cite{Shashkov1995}, so that the resulting diffusion operator is self-adjoint. This results in exact energy conservation and
a vast reduction of unphysical perpendicular heat fluxes. Moreover the order of the scheme is arbitrary. However, the non-aligned discretization suffers from a restrictive Courant-Friedrichs-Lewy (CFL) condition, which is circumvented by an
implicit time integrator. On the other side we implemented two aligned schemes of second order into the dG method, which were motivated from finite differences and rely on 2D interpolation~\cite{Stegmeir2014,Hariri2013}. 
The first scheme is not self-adjoint and fails in conserving energy exactly.
We show numerically that in contrast to the direct aligned FD scheme the convergence is lost for higher resolutions. Hence the direct aligned dG scheme is not robust for general axisymmetric magnetic fields. 
The second aligned scheme is also self-adjoint so that energy is conserved. This scheme is corrected due to jumps at the cell boundaries, which are accompanied by an unfavourable CFL condition and require an implicit treatment. 
Numerical tests reveal irregular and reduced convergence rates, which is in accordance with the results of the self-adjoint aligned FD discretization.
This is owed to corrugations which occur on the grid scale.
However, both aligned dG schemes demonstrate low perpendicular pollution even in cases of coarse resolutions and are particularly advantageous for complex flute-modes with \(k_\parallel \neq 0 \). 
All our proposed schemes are tested for an axisymmetric magnetic field where the divergence of \(\bhat\) is not vanishing.
The computations of the dG schemes were conducted with \textbf{FELTOR} (\textbf{F}ull-F \textbf{EL}ectromagnetic model in \textbf{TOR}oidal geometry),
 which exploits the high degree of parallelism on GPUs and CPUs on shared and distributed memory systems~\cite{Einkemmer2014}.
\textbf{GRILLIX} provides the simulations for the aligned FD schemes via an efficient hybrid parallelization for CPUs~\cite{stegmeir:phd15}.

\section{Theory}\label{sec:theory}
The theory part starts with an introduction to the LDG method in one dimension in Section~\ref{sec:ldgmethod}. 
The generalization to three dimensions in cylindrical coordinates can be used for the non-aligned discretization of the parallel derivative in Section~\ref{sec:parallelna}.
In Section~\ref{sec:parallela} we present the aligned discretizations of the parallel derivative. Those make use of the Legendre polynomials for the 2D interpolation.
The theory part is closed with a short description of the implemented time-discretizations in Section~\ref{sec:timestepping}. 
\subsection{The LDG method} \label{sec:ldgmethod}
In the LDG framework one rewrites the model elliptic equation
\begin{align}\label{ellmod}
 - \frac{\partial }{\partial x }\left(\chi  \frac{\partial f }{\partial x }\right)  &= \rho &\textrm{in}\hspace{2 mm} \Omega 
\end{align}
as a set of first order partial differential equations 
\begin{subequations}
\begin{align}\label{ldg1}
g &=  \frac{\partial f}{\partial x} &\textrm{in}\hspace{2 mm} \Omega \\
q &= \chi g &\textrm{in}\hspace{2 mm} \Omega \\ \label{ldg2}
-\frac{\partial q}{\partial x} &= \rho &\textrm{in} \hspace{2 mm}\Omega
\end{align}
\end{subequations}
with a given function \(\chi(x) >  0  \) on \(\Omega\) and corresponding boundary conditions on \(\partial \Omega\), which are specified later.
In an LDG discretized scheme this set of equations decouples \cite{castillo00}. After a short introduction to Legendre polynomials in Section~\ref{sec:legendre} we 
will derive the discretization of the first Eq.~\eqref{ldg1} in Section~\ref{sec:firstderivatives}. With the help of the adjoint of the first derivative in Section~\ref{sec:adjoint} we obtain a discretization of Eq.~\eqref{ellmod}
in Section~\ref{sec:elliptic}.
In the following we will retain the discussion to one dimension since the generalization to more dimensions is straightforward on orthogonal grids.
\subsubsection{Legendre polynomials}\label{sec:legendre}
Our discontinuous Galerkin methods are based on the use of orthogonal Legendre polynomials up to order $P-1$.
We define 
$x^a_j$ and $w_j$, $j=0,\dots,P-1$ denoting the abscissas and weights of 
the Gauss--Legendre quadrature on the interval $[-1,1]$ and denote
$p_k(x)$ as the $k-th$ Legendre polynomial (see e.g.~\cite{AS}).
With this, we define the forward transformation matrix 
\begin{align}
    F^{kj} &:=\frac{2k+1}{2}w_jp_k(x^a_j).
\end{align}

Let us now consider an interval $[a,b]$ and an equidistant discretization
by $N$ cells with cell center $x_n$ and grid size $h=\frac{b-a}{N}$; in addition, we set $x_{nj}^a := x_n + \frac{h}{2}x^a_j$.
Given a function $f:[a,b]\rightarrow \mathbb{R}$ we then define
$f_{nj} := f(x^a_{nj})$ and denote its dG expansion
\begin{align}
    f_h(x) = \sum_{n=1}^N\sum_{k=0}^{P-1}\bar{f}^{nk} p_{nk}(x)   .
    \label{eq:dgexpansion}
\end{align}
with \(\bar{f}^{nk} :=\sum_{j=0}^{P-1}F^{kj}f_{nj}\). The expansion Eq.~\eqref{eq:dgexpansion} is discontinuous across cell boundaries, hence the name dG.
Here, $p_{nk}(x)$ is the $k-th$ Legendre polynomial in cell $n$
\begin{align}
    p_{nk}(x) := \begin{cases}
        p_k\left(  \frac{2}{h}(x-x_n)\right),& \ \text{for } x-x_n\in\left[ -\frac{h}{2}, \frac{h}{2} \right]\\
        0,& \ \text{else}.
    \end{cases}
    \label{}
\end{align}
Let us define the diagonal weights matrix $W$ and its inverse $V$
\begin{subequations}
\begin{align} \label{weights}
    W^{ij} &:= \frac{h w_j}{2}\delta_{ij}, \\
    V_{ij} &:= W_{ij}^{-1} = \frac{2}{hw_j}\delta_{ij}.
\end{align}
\end{subequations}
The use of Legendre polynomials yields a natural approximation of the $L^2$ scalar product
via Gauss--Legendre quadrature
\begin{subequations}
\begin{align}
    \langle f_h,g_h\rangle:=\int_a^b f_hg_h \dx &= \sum_{n=1}^N\sum_{j=0}^{P-1} \frac{hw_j}{2} f_{nj} g_{nj} =:
		\vec f^{\mathrm{T}}(\Eins\otimes W)\vec g ,\\
    \|f_h\|^2_{L^2} := \int_a^b |f_h|^2 \dx &= \sum_{n=1}^N\sum_{j=0}^{P-1} \frac{h w_j}{2}f_{nj}^2 =:
		\vec f^{\mathrm{T}}(\Eins\otimes W)\vec f,
\label{eq:def_norm}
\end{align}
\label{eq:gausslegendre}
\end{subequations}
where $\otimes$ denotes the Kronecker product and the vector \(\bold{f}\) contains the coefficients \(f_{nj}\). 
With these formulas we have a simple, accurate, and fast 
method to evaluate integrals. This is applied, for example, to compute
errors in the $L^2$-norm.
Function products are easily computed pointwise, i.e.
\begin{align}
    (f_hg_h)_{ni}=f_{ni}g_{ni}.
    \label{}
\end{align}

\subsubsection{The first derivatives} \label{sec:firstderivatives}
The derivative of $p_{ni}(x)$ is not well defined
on cell boundaries, which is why we now retain to 
a weak formulation. 
Consider 
\begin{align}
    \int_{C_n} \partial_x f_h(x) p_{ni}(x) \dx = f_h p_{ni}|_{x_{n-1/2}}^{x_{n+1/2}}  -
    \int_{C_n} f_h(x) \partial_x p_{ni}(x) \dx
    \label{}
\end{align}
on the bounded cell domain \(C_n=\left[x_{n-1/2},x_{n+1/2}\right]\)  on  \(\mathbb{R}\).\\
The approximation $f_h(x)$ is double valued on the cell boundaries. The boundary terms have to be replaced by
\begin{align}
    \int_{C_n} g_h(x) p_{ni}(x) \dx
    = \hat f p_{ni}|_{x_{n-1/2}}^{x_{n+1/2}}  -
    \int_{C_n}\sum_{n=1}^N\sum_{k=0}^{P-1} \bar f^{nk} p_{nk} \partial_x p_{ni}(x) \dx,
    \label{}
\end{align}
where $\hat f(x)$ is the numerical flux across cell boundaries and we call $g_h(x)$ the numerical approximation to the first derivative. 
Let us define 
\begin{subequations}
\begin{align}
    f^+(x) := \lim_{\eps\to 0,\eps>0}f_h(x+\eps),\\
    f^-(x) := \lim_{\eps\to 0,\eps>0}f_h(x-\eps).
    \label{}
\end{align}
\end{subequations}
We construct the numerical flux  by the mean value \(\left\{\left\{ f\right\} \right\} := \left(f^+ + f^-\right)/2\) and the jumps \(\left[ \left[ f \right] \right] := \left(f^+ - f^-\right)/2\) at the cell boundaries 
\begin{align}
 \hat{f} = \left\{\left\{ f\right\} \right\} + C \left[ \left[ f \right] \right]
\end{align}
with \(C:= \left\{0,1,-1\right\} \) determining the centered(C), forward(F) and the backward(B) flux respectively~\cite{Cockburn2001}. 
For $f\colon [a,b]\to\mathbb{R}$ we assume that
\begin{subequations}
\begin{align}
	 f(b+\eps) = f(a+\eps)&,\qquad f(a-\eps) = f(b-\eps), \\
    \hat f(a) &= \hat f(b) = 0,\\
    \hat f(a) = f^+(a) &, \qquad 
    \hat f(b) = f^-(b).
\end{align}
\end{subequations}
for periodic, homogeneous Dirichlet and homogeneous Neumann boundary conditions respectively.
Depending on what flux we choose, we arrive at various 
approximations to the derivative,
e.g. for $P=1$ (i.e. a piecewise constant approximation in each cell)
our scheme reduces to the classic centered, forward and backward finite difference
schemes respectively.

The derivative can be written as a matrix-vector multiplication
\begin{align}
	\mathbf g = D_{x} \mathbf f.
    \label{eq:matrix_xspace}
\end{align}
For a concise notation we define
\begin{align*}
    D_{x} := (\Eins\otimes V)\cdot(\Eins\otimes F^T)\cdot \bar D_{x}\cdot (\Eins\otimes F).
    \label{}
\end{align*}

We now show the matrix representation of the centered, forward, and backward one-dimensional discrete derivative for 
periodic boundary conditions that can be used
in the implementation
\begin{align}
    \bar D^0_{x,per} = \frac{1}{2}\begin{pmatrix}
		(M-M^{\mathrm{T}}) & RL      &    &   & -LR \\
		-LR  & (M-M^{\mathrm{T}}) & RL &   &     \\
             &  -LR    & \dots   &   &     \\
             &         &    & \dots  & RL    \\
			 RL &         &    & -LR&(M-M^{\mathrm{T}}) 
    \end{pmatrix},
    \label{eq:dxcentered}
\end{align}
\begin{align}
    \bar D^+_{x,per} = \begin{pmatrix}
        -(M+LL)^{\mathrm{T}} & RL      &    &   & 0 \\
		 0   & -(M+LL)^{\mathrm{T}} & RL &   &     \\
             &    0   & \dots   &   &     \\
             &         &    & \dots  & RL    \\
			 RL  &         &    & 0 & -(M+LL)^{\mathrm{T}}
    \end{pmatrix},
    \label{eq:dxplus}
\end{align}
and 
\begin{align}
    \bar D^-_{x,per} = \begin{pmatrix}
		(M+LL) & 0      &    &   & -LR \\
		-LR  & (M+LL) & 0 &   &     \\
             &   -LR   & \dots   &   &     \\
             &         &    & \dots  & 0    \\
			 0  &         &    & -LR &(M+LL)
    \end{pmatrix}.
    \label{eq:dxminus}
\end{align}
We used the notation for the (PxP) block-matrices
\begin{subequations}
    \begin{align}
		RR_{ij} &:= 1 = RR^{\mathrm{T}}_{ij}, \quad
        &&RL_{ij}:= (-1)^j,\\
		LL_{ij} &:= (-1)^{i+j} = LL^{\mathrm{T}}_{ij},\quad
		&&LR_{ij}:= (-1)^i = RL^{\mathrm{T}}_{ij},\\
        M_{ij} &:= \begin{cases}
            0 \text{, for }i>j-1\\
            1-(-1)^{i+j} \text{ else}
        \end{cases},
       &&   i,j = (0,\dots P-1),
    \end{align}
    \label{eq:legendre_operators}
\end{subequations}
which we introduce mainly for ease of implementation. If a block-matrix class is written and the
operations $+$, $-$ and $*$ are defined on it, the assembly of the derivative
matrices is simplified to a large extent. 
Note that for $P=1$ we recover the familiar finite difference approximations of the first derivative. 
\begin{table}[htbp]
\caption{Upper left and lower right matrix entries for various boundary conditions. For Dirichlet and von Neumann BC the upper right and lower left entries are zero.}
\scriptsize
\begin{center}
\begin{tabular}{|c|c|c|c|c|c|c|}
\hline
 & \multicolumn{ 2}{c|}{$D_x^+$ (forward)} & \multicolumn{ 2}{c|}{$D_x^-$ (backward)} & \multicolumn{ 2}{c|}{ $D_x^0$ (centered)} \\ \hline
 & left & right & left  & right & left  & right \\ \hline
periodic & $-(M+L)^\mathrm{T}$ & $-(M+L)^\mathrm{T}$ & $(M+L)$ & $(M+L)$ & $\frac{1}{2}(M-M^\mathrm{T})$ & $\frac{1}{2}(M-M^\mathrm{T})$ \\ \hline
Dirichlet & $-M^\mathrm{T}$ & $-(M+L)^\mathrm{T}$ & $(M+L)$ & $-M^\mathrm{T}$ & $\frac{1}{2}(M-M^\mathrm{T}+L)$ & $\frac{1}{2}(M-M^\mathrm{T}-R)$ \\ \hline
Neumann & $-(M+L)^\mathrm{T}$ & $M$ & $M$ & $(M+L)$ & $\frac{1}{2}(M-M^\mathrm{T}-L)$ & $\frac{1}{2}(M-M^\mathrm{T}+R)$ \\ \hline
\end{tabular}
\end{center}
\label{tab:boundary_terms}
\end{table}
Finally, we note the boundary terms for homogeneous Dirichlet and Neumann boundaries
in Table~\eqref{tab:boundary_terms} noticing that only the 
corner entries of the matrices change.
\subsubsection{The adjoint of a matrix}\label{sec:adjoint}
Recall that the adjoint of a matrix $A$ is defined by the scalar product, i.e. 
\begin{align}
    \vec f^\mathrm{T} \cdot\left[(\Eins\otimes W)\cdot A\right]\cdot\vec g = 
    \vec g^\mathrm{T} \cdot\left[A^\mathrm{T}\cdot(1\otimes W)\right]\cdot \vec f =:     \vec g^\mathrm{T}\cdot\left[ (\Eins\otimes W) A^\dagger\right]\cdot \vec f. \nonumber
    \label{}
\end{align}
From here we immediately get the relation
\begin{align}
    A^\dagger \equiv (\Eins\otimes V)\cdot A^\mathrm{T}\cdot (\Eins\otimes W).
    \label{eq:adjoint}
\end{align}
There is a close connection between symmetric, $A=A^\mathrm{T}$, and self-adjoint, $A=A^\dagger$, matrices.
If and only if the matrix $A$ is symmetric, then $(\Eins\otimes V)\cdot A$ is self-adjoint. Of course, we have $(A\cdot B)^\dagger = B^\dagger \cdot A^\dagger$ and $(A^\dagger)^\dagger = A$. 
\subsubsection{Discretization of elliptic equations} \label{sec:elliptic}
We discretize the one-dimensional general elliptic equation Eq.~\eqref{ellmod}
on the interval $\Omega := [a,b]$.
We either choose periodic, Dirichlet, or Neumann boundary conditions on the 
left and right border for $f$. 
Our plan is to simply use one of the discretizations developed in the last 
section for the right derivative 
and its adjoint for the left  derivative 
to obtain an overall selfadjoint scheme. The multiplication with $\chi$ is a simple function product
\begin{align}
  D_x^\dagger \cdot\boldsymbol{\chi}\cdot D_x \vec f  = \boldsymbol{\rho }, 
    \label{eq:naive}
\end{align}
where $D_x$ is either $D_x^+$, $D_x^-$, or $D_x^0$ with the correct boundary terms and $D_x^\dagger$ is its adjoint according to Eq.~\eqref{eq:adjoint}.
Note, that~\cite{Cockburn2001} originally only proposed to use the forward or backward discretization for $D_x$. 
Eq.~\eqref{eq:naive} is indeed a self-adjoint
discretization for the second derivative. However, it turns out that 
Eq.~\eqref{eq:naive} is inconsistent for given $\rho(x)$. The solution does not converge for $P>1$. 
This problem is solved according to~\cite{Cockburn2001} by adding a jump term \(\left[ \left[ f \right] \right]\)
to the numerical flux of $q$ in the case $P>1$, which enhances the stability and accuracy of the LDG method
\begin{align}
 \hat{q} = \left\{\left\{ q\right\} \right\} - C \left[ \left[ q \right] \right] - \left[ \left[ f \right] \right].
\end{align}
That means we have to alter our discretization~\eqref{eq:naive} according to
\begin{align}
     \left[D_x^\dagger\cdot  \boldsymbol{\chi}\cdot D_x  + J\right]\vec f  =  \boldsymbol{\rho},
    \label{eq:discreteelliptic}
\end{align}
where again $J=(\Eins\otimes V)(\Eins\otimes F^\mathrm{T}) \bar J (\Eins\otimes F)$ and
\begin{align}
    \bar J = \begin{pmatrix}
		(LL+RR) & -RL      &    &   & -LR \\
		-LR  & (LL+RR) & -RL &   &     \\
             &   -LR   & \dots   &   &     \\
             &         &    & \dots  & -RL    \\
			 -RL  &         &    & -LR &(LL+RR)
    \end{pmatrix}
\end{align}
for periodic boundaries. Again we give the correct boundary terms for Dirichlet
and Neumann boundary conditions in Table~\ref{tab:jump_terms}.
\begin{table}[htbp]
\begin{center}
\begin{tabular}{|c|c|c|}
\hline
 & \multicolumn{ 2}{c|}{$J$} \\ \hline
 & left & right \\ \hline
periodic & $L+R$ & $L+R$ \\ \hline
Dirichlet & $L+R$ & $L+R$ \\ \hline
Neumann & $R$ & $L$ \\ \hline
\end{tabular}
\end{center}
\caption{Top left and bottom right entries for jump matrix}
\label{tab:jump_terms}
\end{table}
Note that $\bar J$ is symmetric, thus the overall discretization remains self-adjoint. Indeed, with $D_x = D_x^+$ Eq.~\eqref{eq:discreteelliptic} recovers
the discretization proposed by~\cite{Cockburn2001}. 
In addition, we remark that the centered discretization in Eq.~\eqref{eq:discreteelliptic} is symmetric with respect to an inversion of the coordinate system $x\rightarrow -x$ even for double Dirichlet or Neumann boundaries, 
while the forward and backward discretization is not. 
\subsection{Non-aligned discretization of the parallel derivative} \label{sec:parallelna}
The generalization of the LDG method to higher dimensions for orthogonal grids is straightforward. 
All the matrices derived above can readily be extended via the appropriate Kronecker products.
The space complexity of the matrices derived is $\mathcal{O}(P^2 N)$ in one, $\mathcal{O}(P^3N^2)$ in two and $\mathcal{O}({P^4 N^3})$ in three dimensions. 
In our case we use cylindrical coordinates \(\{R,Z,\varphi\}\), so that the
volume form reads $R\d R\d Z\d \varphi$. 
All our presented dG methods are discretized on 
a cylindrical mesh, which  consistis of \(N_R N_Z N_\varphi\) cells with \(P_R P_Z O \varphi\) polynomial coefficients.
This has to be taken into account by multiplying the weights in Eq.~\eqref{eq:adjoint} by the factor $R$ when computing the adjoint of matrices.
We do so for the parallel derivative, which can be written as  $\nabla_\parallel =b^R\partial_R + b^Z\partial_Z + b^\varphi\partial_\varphi$. The discrete centered, forward and backward derivatives are 
\begin{align}
 {G}_0 = b^R \circ D_R^0 + b^Z \circ D_Z^0 + b^\varphi \circ D_\varphi^0,\\
 {G}_+ = b^R \circ D_R^+ + b^Z \circ D_Z^+ + b^\varphi \circ D_\varphi^+,\\
 {G}_- = b^R \circ D_R^- + b^Z \circ D_Z^- + b^\varphi \circ D_\varphi^-,
\end{align}
where \(\circ\) denotes pointwise multiplication.
\subsection{Aligned discretization of the parallel derivative} \label{sec:parallela}
We begin with the formulation of a field-aligned discretization. If $s$ denotes the 
field line following coordinate, then the one-dimensional discrete derivative along the field line reads
\begin{align}
    \frac{\d f }{\d s} \rightarrow \frac{f_{k+1}-f_{k-1}}{s_{k+1} - s_{k-1}}.
    \label{eq:proposition}
\end{align}
From differential geometry we 
know that to every smooth vector field $\bhat$ there is a unique curve of which the
tangent in a point is the value of $\bhat$ at that point\cite{Frankel}. It is given by
the solution of the differential equation
\begin{align}
    \frac{\d z^i}{\d s} = \hat b^i(\vec z),
    \label{eq:integralcurve}
\end{align}
where $z^i$ is one of $(R, Z, \varphi)$ and $\hat b^i$ are the contravariant components
of $\bhat$ in cylindrical coordinates. 
Moreover, by definition we have
\begin{align}
    \frac{\d f(\vec z(s))}{\d s} = \bhat\cdot \nabla f|_{\vec z(s)}
    \label{}
\end{align}
along a field line parameterized by distance $s$, 
i.e. instead of $\nabla_\parallel f$ we can choose to discretize $\frac{\d f}{\d s}$.

Let us divide the $\varphi$ direction into $N_\varphi$ equidistant planes of distance
$\Delta \varphi$. Unfortunately, from Eq.~\eqref{eq:integralcurve} we cannot easily determine the 
distance $\Delta s$ for given $\Delta \varphi$. It is better to integrate
\begin{align}
    \frac{\d z^i}{\d t}=\frac{b^i}{b^\varphi} = \frac{B^i}{B^\varphi}
    \label{}
\end{align}
since in this case $\d\varphi/\d t = 1 \Rightarrow t=\varphi$. We get
\begin{subequations}
\begin{align}
    \frac{\d R}{\d\varphi}&= \frac{B^R}{B^\varphi},\\ 
    \frac{\d Z}{\d\varphi}&=\frac{B^Z}{B^\varphi},
\end{align}
\text{ together with the equation  }
\begin{align}
    \frac{\d s}{\d\varphi} &= \frac{1}{|\hat b^\varphi|} = \frac{B}{|B^\varphi|}
    \label{eq:fieldlinec}
\end{align}
\label{eq:fieldline}
\end{subequations}
for the length of the field line $s$. 
Eqs.~\eqref{eq:fieldline} are integrated from $\varphi=0$ to $\varphi=\pm \Delta \varphi$. 
We characterize the flow generated by $\bhat/b^\varphi$ by
\begin{align}
    T_{\Delta \varphi}^{\pm 1}\vec z := T_{\Delta \varphi}^{\pm 1}[R, Z, \varphi]:= ( R(\pm \Delta\varphi), Z( \pm \Delta\varphi), \varphi\pm\Delta \varphi),
    \label{}
\end{align}
where $(R(\varphi), Z(\varphi), s(\varphi))$ is the solution to Eqs.~\eqref{eq:fieldline} 
with initial condition 
\begin{align}
    (R(0), Z(0), s(0)) = (R, Z, 0).
    \label{}
\end{align} 
Obviously we have $T^{-1}_{\Delta\varphi}\cdot T^{+1}_{\Delta\varphi} = \Eins$, but $T^{\pm}_{\Delta\varphi}$ is not unitary since $\bhat/b^\varphi$ is 
not divergence free. 

The proposed centered discretization~\eqref{eq:proposition} for the parallel derivative then reads
\begin{align}
    \nabla_\parallel f \equiv \frac{df}{ds} = \frac{df}{d\varphi}\frac{d\varphi}{ds} 
    \rightarrow \frac{f\left(T_{\Delta\varphi}^+\vec z\right)-f\left(T_{\Delta\varphi}^-\vec z\right)}{s(+\Delta\varphi) - s(-\Delta\varphi)},
    \label{eq:paralleldis}
\end{align}
which is slightly different from Reference~\cite{Hariri2014}, where
the relation~\eqref{eq:fieldlinec} was used to replace $\d \varphi/\d s$.
Previous derivations needed to construct an interpolation scheme in order
to evaluate functions on the transformed coordinates, which in general do not coincide with grid points.
Note that \(f\) is given as a dG expansion in \(R,Z,\varphi\) with \(P_{R}\),\(P_{Z}\) arbitrary and \(P_{\varphi}=1\).
Hence the interpolation of the transformed points $T_{\Delta\varphi}^{\pm 1}\vec z$ is naturally given by Eq. (7) (the extension to three dimensions is immediate). 
If $(R_{ni},Z_{mj},\varphi_k)$ are the grid points, 
we call $(R^+_{ni}, Z^+_{mj},\varphi_{k+1}) := T^+_{\Delta\varphi}[R_{ni},Z_{mj}, \varphi_k]$ and $(R_{ni}^-, Z^-_{mj},\varphi_{k-1}) := T^-_{\Delta\varphi}[R_{ni},Z_{mj}, \varphi_k]$ the transformed coordinates along
the field lines. We then have
\begin{subequations}
\begin{align}
    f(T^+_{\Delta\varphi}\vec z) = f( R^+_{ni},Z^+_{mj}, \varphi_{k+1}) &= \sum_{m_R=1}^{N_R} \sum_{m_Z=1}^{N_Z} \sum_{l_R=0}^{P_R-1}\sum_{l_Z=0}^{P_Z-1} \bar
                    f_{k+1}^{m_R l_R m_Z l_Z}p_{m_R l_R }(R^+_{ni})p_{m_Z l_Z}(Z^+_{mj})  \nonumber \\ &
=:  \sum_{m_\varphi=1}^{N_\varphi} \sum_{m_R=1}^{N_R} \sum_{m_Z=1}^{N_Z}  \sum_{j_R=0}^{P_R-1}\sum_{j_Z=0}^{P_Z-1} 
                    (I^+)_{knimj}^{m_\varphi m_R  j_R m_Z  j_Z}f_{m_\varphi m_R j_R m_Z j_Z} , \\
    f(T^-_{\Delta\varphi}\vec z) = f( R^-_{ni},Z^-  _{mj},\varphi_{k-1}) &=\sum_{m_R=1}^{N_R} \sum_{m_Z=1}^{N_Z} \sum_{l_R=0}^{P_R-1}\sum_{l_Z=0}^{P_Z-1}  \bar 
    f_{k-1}^{m_R l_R m_Z l_Z}p_{m_R l_R}(R^-_{ni}) p_{m_Z l_Z}(Z^-_{mj})   \nonumber \\ &
=: \sum_{m_\varphi=1}^{N_\varphi} \sum_{m_R=1}^{N_R} \sum_{m_Z=1}^{N_Z} \sum_{j_R=0}^{P_R-1}\sum_{j_Z=0}^{P_Z-1}   (I^-)_{knimj}^{m_\varphi m_R  j_R m_Z  j_Z }f_{m_\varphi m_R j_R m_Z j_Z} , 
\end{align}
\end{subequations}
where the backward transformations $\bar{ \vec f}:= (\Eins\otimes F) \vec{f}$ are hidden in $I$.
Thus, the interpolation of all the necessary points can simply be written as a matrix-vector product, where the interpolation matrices $I^+$  and $I^-$ are independent of time since
the magnetic field is assumed constant in time. The order of this interpolation is given by $P$, the number of polynomial coefficients.
A consistency check is the relation $I^+\cdot I^- = \Eins$. 

The centered discretization~\eqref{eq:paralleldis} can now be written as a matrix vector product according to
\begin{align}
{Q}_{0}f&= S_0 \cdot\left[I^+ - I^- \right] f,
\end{align}
where $S_0$ is the diagonal matrix that contains the entries $\left(s(+\Delta\varphi) - s(-\Delta\varphi)\right)^{-1}$.
This discretization is not skew-symmetric since the
field lines are not volume-preserving, or~$(I^+)^\mathrm{T} \neq I^-$.  The forward  \((+)\) and backward  \((-)\) operators \({Q} \) for the parallel derivative are formulated analogously to
\begin{subequations}
\begin{align}
{Q}_{+} f&= S_+ \cdot \left[I^+ - I \right] f, \\
{Q}_{-}f&= S_- \cdot \left[I^- - I \right] f, 
\end{align}
\end{subequations}
where $S_\pm$ is the diagonal matrix that contains the entries $\left(s(\pm\Delta\varphi) \right)^{-1}$.
\subsubsection{Boundary conditions}\label{sec:bc}
The main problem with the above scheme is the question what  to do when a field line crosses the simulation boundaries. 
Boundary conditions(BC) are formulated in cylindrical coordinates and are either homogeneous Dirichlet or Neumann. The latter 
are straightforward and implemented by setting \(\nabla_\parallel f = 0\) if the field line leaves the computational domain. For Dirichlet BCs
one idea is to simply cut the contribution from field lines
that leave the computational domain. While this works in practice
it is unclear what numerical and physical side-effects this procedure might have. 
Another idea is to check to every point $\vec z$ whether $T_{\Delta\varphi}\vec z$
lies inside our simulation box or not. If not, we have to find where exactly the 
field line intersects the simulation box.  We have to find
$\varphi_b$ such that the result of the integration of Eq.~\eqref{eq:fieldline} from 
$0$ to $\varphi_b$ lies on the boundary. 
The angle $\varphi_b$ can be found by a bisection algorithm knowing that $0<\varphi_b < \Delta\varphi$. 
This kind of procedure is known as a shooting method. 
When all points are found, ghost cells can be constructed in the correct way. 

\subsection{Time-stepping}\label{sec:timestepping}
Having discretized the spatial derivatives of Eq.~\eqref{eq:diffusion}, 
we can now independently choose a discretization in time.
This is known as the method of lines. In order to be able to test both explicit and implicit methods
we implemented a semi-implicit Runge-Kutta (SIRK) scheme~\cite{Yoh2004}. 
The scheme is of third order
in the explicit terms and of second order in the implicit terms. 
The method combines the cost-effectiveness of explicit algorithms
with the stability of purely implicit time-steppers. We solve the implicit substeps by a conjugate gradient method, which works as long 
as the implicit part remains symmetric and linear. 
We finally remark that, given the explicit and implicit parts, an implementation of the time-stepper only require linear algebra vector space operations. 
In fact, all of the up to now discussed algorithms can be implemented using basic vector-vector 
operations and a sparse matrix-vector product.

\section{Three dG schemes}\label{sec:three}
We are now ready to propose three different numerical approximations \({D}\) to the parallel Diffusion operator \(\Delta_\parallel\). 
\subsection{Direct aligned dG discretization} \label{sec:threeDirect}
In the first approximation we apply the divergence free property of the magnetic field $\vec B$ and write $\nabla\cdot( \bhat f)=B\nabla_\parallel \left(B^{-1} f\right)$. 
We then use one of the centered, forward or backward aligned discretizations of the parallel derivative of Section~\ref{sec:parallela}
to get a discretization of $\Delta_\parallel f = B\nabla_\parallel \left(B^{-1} \nabla_\parallel f\right)$. 
Hence, we can formulate two schemes based on either centered differences or an arithmetic average over the forward and backward differences
\begin{align}
    {D}_{0}^D f &=B {Q}_{0} B^{-1} {Q}_{0}f, & {D}_{a}^D f &= \frac{1}{2}\left[ B {Q}_{+} B^{-1} {Q}_{+}f + B {Q}_{-} B^{-1}  {Q}_{-}f \right].
\end{align}
In contrast to the naive discretization of Reference~\cite{Stegmeir2014},
this scheme holds for magnetic fields with \(\vec{\nabla} \cdot \vec{b} \neq 0\) and makes use of the Legendre polynomials for the interpolation.
\subsection{Self-adjoint aligned dG discretization} \label{sec:threeSelf}
We note the adjoint of the parallel derivative:
\begin{align}
    \nabla_\parallel^\dagger f = - \nabla\cdot\left(\bhat\ f \right) .
    \label{}
\end{align}
With this relation we can define the parallel diffusion operator of Eq.~\eqref{eq:definition} to be self-adjoint according to
\begin{align}
    \Delta_\parallel = -\nabla_\parallel^\dagger \nabla_\parallel .
    \label{}
\end{align}
This leads to the idea to simply use a discretization for the 
parallel derivative and use the matrix adjoint Eq.~\eqref{eq:adjoint} to get a discretization for the divergence.
Hence, we can derive the self-adjoint discretization for centered and averaged differences
\begin{align}
 {D}_{0}^A f &=\left[{Q}^\dagger_{0}  {Q}^{}_{0}+ J \right]f,
 & {D}_{a}^A  f &= \frac{1}{2}\left[{Q}_{+}^\dagger  {Q}_{+}^{} +{Q}_{-}^\dagger  {Q}_{-}^{}  +2  J \right] f.
\end{align}
The scheme is similar to that of Reference~\cite{Stegmeir2014} except for the interpolation technique, and that in the dG framework 
we have to add the jump term J of Eq.~\eqref{eq:discreteelliptic} to obtain a convergent and stable diffusion operator.
\subsection{Self-adjoint non-aligned LDG discretization}\label{sec:threeNon}
The directional derivative $\nabla_\parallel = \bhat\cdot\nabla$ can be discretized directly as $\nabla_\parallel =b^R\partial_R + b^Z\partial_Z + b^\varphi\partial_\varphi$ using the dG methods developed in the 
Section~\ref{sec:firstderivatives}. The adjoint of the resulting matrix is
then a discretization for the divergence $\nabla\cdot(\bhat f)$ again. The compostion of both yields then a discretization for the parallel Laplacian according to
\begin{align}
 {D}_{0}^{G} f &=\left[{G}_{0}^\dagger  {G}_{0}^{} + J \right]f, & 
 {D}_{a}^{G} f &= \frac{1}{2}\left[{G}_{+}^\dagger  {G}_{+}^{} +{G}_{-}^\dagger  {G}_{-}^{} + 2 J \right]f.
\end{align}
Note, that we add jump terms for stabilization once more.
\section{Two aligned FD schemes}\label{sec:threeFD}
The dG schemes are compared to two aligned FD discretizations, presented in References~\cite{Stegmeir2014,Stegmeir2015,stegmeir:phd15}. 
The first aligned FD scheme is a generalization of the naive scheme, which is based on third order bipolynomial interpolation (\textbf{N-3}) and
was originally proposed to discretize $\nabla_\parallel^2$. 
The operator \(\left(\vec{\nabla} \cdot \bhat\right) \nabla_\parallel\) was added to this scheme so that effects arising from \(\vec{\nabla} \cdot \bhat\neq0\) are taken into account.
The discrete parallel gradient is thereby computed via a central finite difference along magnetic field lines~\cite{Hariri2013},
and the resultant scheme is denominated \textbf{D-3}.
The second aligned FD schemes \textbf{S-3} is formulated in the spirit of the SOM. In the \textbf{S-3} scheme the discrete parallel 
gradient is also computed with the help of third order bipolynomial interpolation.
The parallel diffusion operator then follows via requiring its self-adjointness property on the discrete level.
 
\section{Tests}\label{sec:numericalexperiments}
We use an axisymmetric magnetic field with \(\vec{\nabla} \cdot \bhat \neq 0\). In cylindrical coordinates $\{R,Z,\varphi\}$ it is given by
\begin{align}
 \vec{B} = I_0 \vec{\nabla} \varphi + \vec{\nabla} \psi_p \times \vec{\nabla} \varphi,
\end{align}
with $I_0=20$, $R_0=10$ and a poloidal flux of the form
\begin{align}
 \psi_p &= \cos{\left(\frac{\pi}{2}(R-R_0)\right)} \cos{\left(\frac{\pi}{2} Z\right) }.
\end{align}
We perform all our numerical tests on the domain \(R\in[9,11]\), \(Z\in[-1,1]\) and \(\varphi\in[0,2 \pi]\), except for the test given in Section~\ref{sec:tokamak}. 
With this choice of parameters the magnetic field never crosses the simulation box.
\subsection{Convergence test to an analytical test function}\label{sec:anaconvtestdg}
First, we define a general test function \(f = -\psi_p(R,Z)\cos{(\varphi)} \), which fulfills Dirichlet as well as Neumann boundary conditions. 
We use this function to directly compute the relative error \( \epsilon_{\Delta_\parallel f} =  \|{D} f_{} -\Delta_\parallel f \|_{L^2}  / \| \Delta_\parallel f \|_{L^2}\) 
between the numerical diffusion operator \({D} f\) and the analytical solution \(\Delta_\parallel f\). 
At this point we introduce also a second relative error \( \epsilon_{ f} =  \|  f_{{D}} - f \|_{L^2} / \|f \|_{L^2} \), which is computed by inverting the diffusion operator with a conjugate gradient method. 
On the continuous level this would result in an ill-posed problem. However, the discrete parallel diffusion operator contains always a small amount of numerical perpendicular diffusion, which makes the discrete operator invertible.\\
All the so far mentioned aligned schemes are constructed to converge with order two in \(\Delta \varphi\) for a homogeneous magnetic field~\cite{Stegmeir2015}. Moreover, the order of the non-aligned scheme is arbitrary.
We define this as the optimal order. 
For inhomogeneous magnetic fields we can observe this optimal order only if we increase the resolution in \(N_R\), \(N_Z\) and \(N_\varphi\) with the correct factors, which depend on the
interpolation order and the order in \(\varphi\)-direction.
Our tested schemes are limited to  \(P_{RZ} = 3\) and  \(P_{\varphi}=1 \). 
For the non-aligned LDG method we additionally investigate the convergence rates for  \(P_{RZ} = 3\) and  \(P_{\varphi}=3 \). \\

\subsubsection{dG schemes}\label{sec:convdgschem}
We are now ready to show the orders of convergence of the centered and averaged discretizations. For this purpose we increase the resolutions in all directions with the correct order according to 
\(P_{RZ}=3\), \(N_R=N_Z=5\hspace{1 mm} (2^{i \hspace{0.5 mm} 2 /3} )\), \(N_\varphi=5 \hspace{1 mm} (2^i)\) with \(i=\left\{0,1,\dots,5\right\} \). The relative errors and orders of convergence are depicted  in 
Table~\ref{table:centeredconv} for the centered schemes and Table~\ref{table:averagedconv} for the averaged schemes. 
\begin{table}[!ht]
\begin{center}
\caption{\(L^2\) error and orders of accuracy (\( \textrm{order}_{\Delta \phi}:= \log(L^2 \textrm{error}_{{\Delta \phi}^{-}}/L^2 \textrm{error}_{\Delta \phi})/\log({\Delta \phi}^{-}/{\Delta \phi}) \)) for the centered non-aligned LDG scheme, 
the centered direct aligned scheme and the centered self-adjoint aligned scheme. The \(L^2\) error for the self-adjoint schemes is
computed with  \(\epsilon_{ f} \) and for the direct aligned scheme with \(\epsilon_{\Delta_\parallel f}\)}
\begin{tabular}{||c||c c||c c||c c|||c||c c||}
\hline
\multicolumn{1}{||l||}{} &  \multicolumn{6}{c|||}{\(P_\varphi=1\)} &  \multicolumn{1}{l||}{} & \multicolumn{2}{c||}{\(P_\varphi=3\)} \\ \hline
\multicolumn{1}{||l||}{} & \multicolumn{2}{c||}{ \(D^G_0\) } &  \multicolumn{2}{c||}{\(D^D_0\)} &   \multicolumn{2}{c|||}{\(D^A_0\)} & \multicolumn{1}{l||}{} & \multicolumn{2}{c||}{\(D^G_0\)} \\ 
 $\Delta \varphi$  &  $L^2$ error  & order &  $L^2$ error  & order &  $L^2$ error  & order & $\Delta \varphi$  &  $L^2$ error  & order \\ \hline
\(2 \pi/3\)& 4.85E+00 &  -- & 8.29E-01 &  -- & 6.65E+00 & -- & \(2 \pi/3\) & 3.88E-02 & --  \\ 
\(2 \pi/5\) & 7.46E-01 & 3.66 & 4.28E-01 & 1.30 & 1.11E+00 & 3.51 & \(2 \pi/5\)& 7.58E-03 & 3.20  \\ 
\(2 \pi/10\) & 1.43E-01 & 2.39 & 1.25E-01 & 1.77 & 3.39E-01 & 1.71 & \(2 \pi/10\)& 7.95E-04 & 3.25 \\
\(2 \pi/20\)& 3.36E-02 & 2.09 & 3.26E-02 & 1.94 & 1.92E-01 & 0.82  & \(2 \pi/20\)& 6.77E-05 & 3.55\\ 
\(2 \pi/40\)& 8.27E-03 & 2.02 & 8.24E-03 & 1.98 & 1.25E-01 & 0.62  & \(2 \pi/30\)& 1.61E-05 & 3.55\\ 
\(2 \pi/80\)& 2.08E-03 & 1.99 & 2.09E-03 & 1.98 & 4.50E-02 & 1.48 &-- & -- & --\\ 
\(2 \pi/160\)& 5.14E-04 & 2.02 & 5.94E-04 & 1.82 & 6.00E-02 & -0.42 &-- & -- & --\\ \hline
\end{tabular}
\label{table:centeredconv}
\end{center}
\end{table}
\begin{table}[!ht]
\begin{center}
\caption{\(L^2\) error and orders of accuracy (\( \textrm{order}_{\Delta \phi}:= \log(L^2 \textrm{error}_{{\Delta \phi}^{-}}/L^2 \textrm{error}_{\Delta \phi})/\log({\Delta \phi}^{-}/{\Delta \phi}) \)) for the averaged non-aligned LDG scheme, the averaged direct aligned scheme and the averaged aligned self-adjoint scheme.
The \(L^2\) error for the self-adjoint schemes is
computed with  \(\epsilon_{ f} \) and for the direct aligned scheme with \(\epsilon_{\Delta_\parallel f}\)}
\begin{tabular}{||c||c c||c c||c c|||c||c c||}
\hline
\multicolumn{1}{||l||}{} &  \multicolumn{6}{c|||}{\(P_\varphi=1\)} &  \multicolumn{1}{l||}{} & \multicolumn{2}{c||}{\(P_\varphi=3\)} \\ \hline
\multicolumn{1}{||l||}{} & \multicolumn{2}{c||}{ \(D^G_a\) } &  \multicolumn{2}{c||}{\(D^D_a\)} &   \multicolumn{2}{c|||}{\(D^A_a\)} & \multicolumn{1}{l||}{} & \multicolumn{2}{c||}{\(D^G_a\) } \\
 $\Delta \varphi$ & $L^2$ error & order & $L^2$ error & order & $L^2$ error & order &  $\Delta \varphi$  &  $L^2$ error  & order \\ \hline
\(2 \pi/3\) & 4.62E-01 &  -- & 3.17E-01 & -- & 7.77E-01 &  -- & \(2 \pi/3\) & 3.95E-02 &--\\ 
\(2 \pi/5\)& 1.43E-01 & 2.30 & 1.25E-01 & 1.82 & 2.13E-01 & 2.53 & \(2 \pi/5\)& 7.96E-03 & 3.14  \\
\(2 \pi/10\)& 3.36E-02 & 2.09 & 3.26E-02 & 1.94 & 6.66E-02 & 1.68 & \(2 \pi/10\) & 8.46E-04 & 3.23  \\
\(2 \pi/20\)& 8.27E-03 & 2.02 & 8.24E-03 & 1.98 & 4.43E-02 & 0.59 & \(2 \pi/20\) & 8.94E-05 & 3.24  \\
\(2 \pi/40\)& 2.06E-03 & 2.01 & 2.17E-03 & 1.92 & 2.42E-02 & 0.87 & \(2 \pi/30\) & 2.42E-05 & 3.22 \\
\(2 \pi/80\)& 5.15E-04 & 2.00 & 7.69E-04 & 1.50 & 2.29E-02 & 0.08 & -- & -- & -- \\
\(2 \pi/160\)& 1.29E-04 & 2.00 & 5.05E-04 & 0.61 & 4.46E-02 & -0.96 &-- & -- & -- \\ \hline
\end{tabular}
\label{table:averagedconv}
\end{center}
\end{table}
Those tables reveal that the non-aligned LDG discretization is the only scheme, which has
the optimal order of convergence for \(P_\varphi = 1\) and also superconvergent rates for \(P_\varphi = 3\). 
The aligned schemes do both show up stagnating convergence rates for higher resolutions whereas the self-adjoint aligned scheme additionally 
fails in converging with the optimal order of two. Instead it converges with a reduced and irregular order, casting it into a true supraconvergent discretization.
This is a consequence of oscillatory relative error fields, which are shown in Figure~\ref{fig:diff} and discussed throughout the next sections.
Furthermore, we note that we observed stagnating convergence rates also for the relative error of the parallel gradient operator \({Q}_{\left\{0,+,-\right\}}\). 
\subsubsection{Aligned FD schemes}\label{sec:convfdschem}
\begin{figure}[!ht]
\centering
\hspace{0.13\textwidth} \(D^A_a\)\hspace{0.35\textwidth}\textbf{S-3}\newline
   \includegraphics[trim = 0px 0px 0px 0px, clip, scale=0.65]{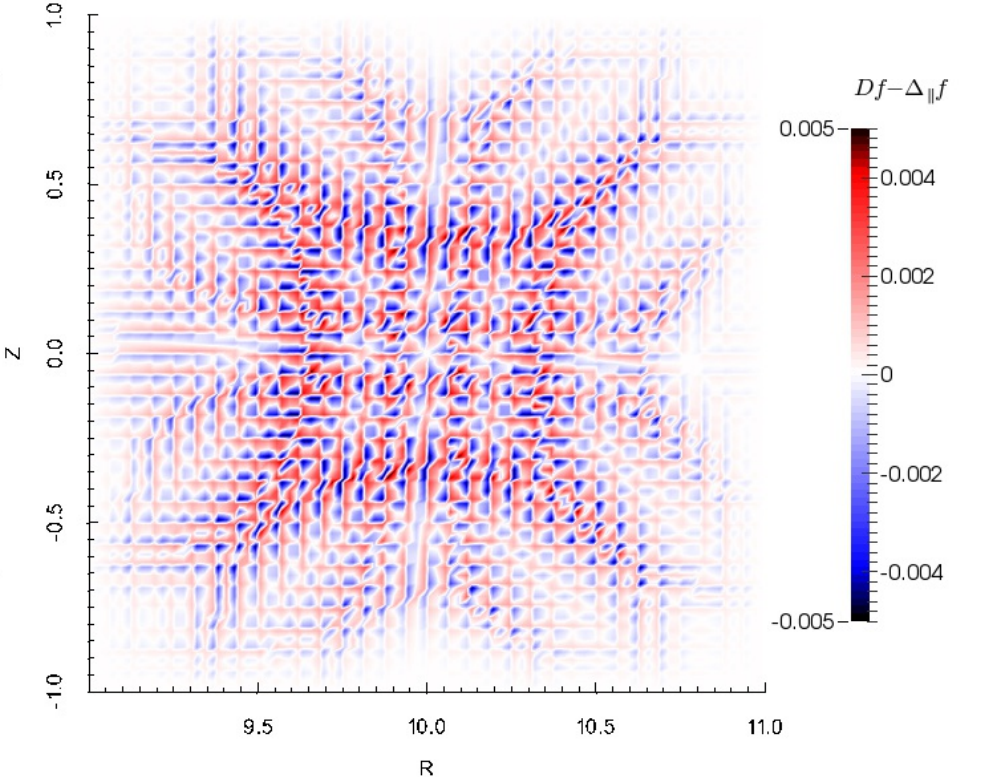}
   \includegraphics[trim = 0px 0px 0px 0px, clip, scale=0.65]{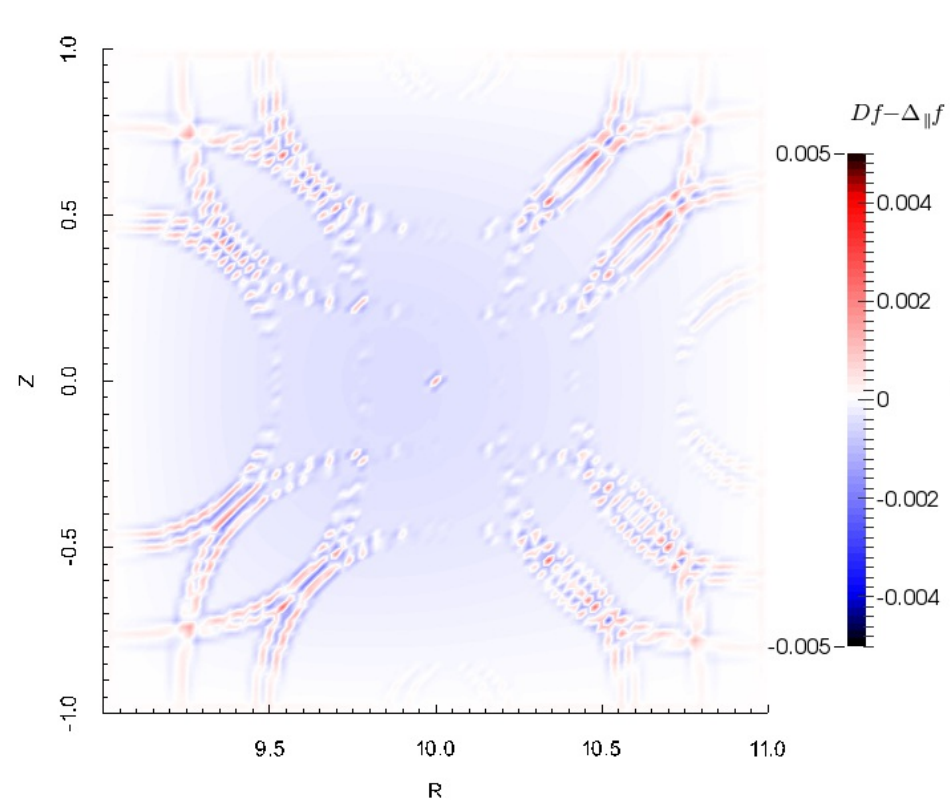}
\caption{The error fields \({D} f_{} -\Delta_\parallel f\) of the convergence test are shown for the self-adjoint aligned dG scheme \(D^A_a\) and the self-adjoint aligned FD scheme \textbf{S-3} at \(\varphi = 0\).
The resolution is \( P_{\varphi}=1\), \( P_{RZ}=3\), \(N_\varphi= 80\), \(N_R=N_Z=32\) and equally \(N_\varphi= 80\),\(N_R=N_Z=100\) for the latter. In contrast to the \textbf{S-3} scheme the oscillations/corrugations 
are more pronounced for the \(D^A_a\) scheme.}
\label{fig:diff}
\end{figure}

Additionally the same convergence test was performed with the \textbf{GRILLIX} code, which is based on the aligned FD schemes depicted in Section~\ref{sec:threeFD}.
\\
In Table~\ref{table_conv_varperp} the relative error $\epsilon_{\Delta_\parallel f}$ and convergence rates for the aligned FD schemes 
are shown. We adapted the resolution according to roughly $N_R=N_Z=3\cdot 5(2^{i2/3})$, $N_\varphi=5(2^i)$, $i=\left\{0,1,\dots6\right\}$, 
where the additional factor of $3$ at $N_R,N_Z$ shall compensate for the factor $P_{RZ}=3$ used in the test of Section~\ref{sec:convdgschem}.
The derived relative error for the \textbf{D-3} 
scheme is of order two.
However, similar to the $D_a^D$ scheme it does not conserve the self-adjointness property of the parallel diffusion operator and therefore does not conserve the $L^1$ norm, 
i.e.~energy (see also Section~\ref{sec:Gaussian}).
The \textbf{S-3} scheme exhibits an irregular and slower convergence behaviour. This is owed to corrugations (see Figure~\ref{fig:diff}), 
which occur if the transformed coordinates $\mathbf{T}^\pm_{\Delta\varphi}\left[R_{i},Z_{j},\varphi_k\right]$ of two neighbouring grid points are in the same grid cell, 
respectively jump across grid cells. The mechanism is described in more detail in  Reference~\cite{Stegmeir2015}. 
Apart from these corrugations the error is of the same level as for the \textbf{D-3} scheme. In general, also for magnetic fields with 
\(\vec{\nabla} \cdot \bhat=0\) corrugations can occur and have been observed, but they are more pronounced in situations with \(\vec{\nabla} \cdot \bhat\neq0\).
As long as the corrugations remain on the grid scale (a suitable criterion for this is given in Reference~\cite{Stegmeir2015}) they could possibly cured by adding a small amount of perpendicular diffusion, 
which is usually anyway present in turbulence simulations. 
\begin{table}[ht]
\begin{center}
\caption{Relative error $\epsilon_{\Delta_\parallel f}$ and its order of accuracy (\( \textrm{order}_{\Delta \phi}:= \log(L^2 \textrm{error}_{{\Delta \phi}^{-}}/L^2 \textrm{error}_{\Delta \phi})/\log({\Delta \phi}^{-}/{\Delta \phi}) \)) 
for the aligned FD schemes.}
\begin{tabular}{||c|c||cc||cc||}
\hline
 &  & \multicolumn{2}{c||}{\textbf{D-3}} & \multicolumn{2}{c||}{\textbf{S-3}}  \\
 $\Delta\varphi$ & $N_{R,Z}\, (h)$ & $L^2$ error & order & $L^2$ error & order  \\ \hline
 $2\pi/3$  & $20$  (\esc{1.00}{-}{01}) & \esc{3.06}{-}{01} & $-$ & \esc{3.06}{-}{01} & $-$   \\
 $2\pi/5$  & $20$  (\esc{1.00}{-}{01}) & \esc{1.20}{-}{01} & $1.81$ & \esc{1.42}{-}{01} & $1.49$  \\   
 $2\pi/10$ & $24$  (\esc{8.33}{-}{02}) & \esc{3.28}{-}{02} & $1.89$ & \esc{1.16}{-}{01} & $0.30$  \\     
 $2\pi/20$ & $40$  (\esc{5.00}{-}{02}) & \esc{5.12}{-}{03} & $2.00$ & \esc{4.42}{-}{02} & $1.39$ \\
 $2\pi/40$ & $60$  (\esc{3.33}{-}{02}) & \esc{2.12}{-}{03} & $1.95$ & \esc{2.49}{-}{02} & $0.83$  \\     
 $2\pi/80$ & $100$ (\esc{2.00}{-}{02}) & \esc{5.12}{-}{04} & $2.04$ & \esc{1.65}{-}{02} & $0.59$  \\     
 $2\pi/160$& $150$ (\esc{1.33}{-}{02}) & \esc{1.25}{-}{04} & $2.04$ & \esc{1.24}{-}{02} & $0.41$  \\     
 $2\pi/320$& $240$ (\esc{8.33}{-}{03}) & \esc{2.81}{-}{05} & $2.14$ & \esc{9.35}{-}{03} & $0.41$  \\     
\hline
\end{tabular}
\label{table_conv_varperp}
\end{center}
\end{table}
\subsection{Numerical simulations} \label{sec:simulations}
We investigate now the conservation properties and the convergence behavior of our presented schemes in numerical simulations of Eq.~\eqref{eq:diffusion}. 
In this respect this we use Neumann boundary conditions to show explicitly the proposed energy 
conservation of our self-adjoint schemes. The convergence is examined
for a general non-aligned three dimensional Gaussian blob in Section~\ref{sec:Gaussian}. The maintenance of the most basic \(k_\parallel = 0\) mode is shown for all schemes in Section~\ref{sec:simulations}.
We take \(\chi_\parallel = 100\) and \(\chi_\perp = 0\) for all of our simulations. The time integration was either performed with a second-order implicit or third-order explicit Runge-Kutta scheme. 
The time step was fixed to \(\Delta t=0.001 \) so that the time error can be ignored. The convergence test was performed with an even smaller timestep.
\subsubsection{Gaussian Blob} \label{sec:Gaussian}
\begin{figure}[!ht]
\centering
\hspace{0.08\textwidth}\(D^G_a\)\hspace{0.31\textwidth}\(D^D_a\)\hspace{0.31\textwidth}\(D^A_a\)\newline
    \includegraphics[trim = 0px 0px 0px 0px, clip, scale=0.8]{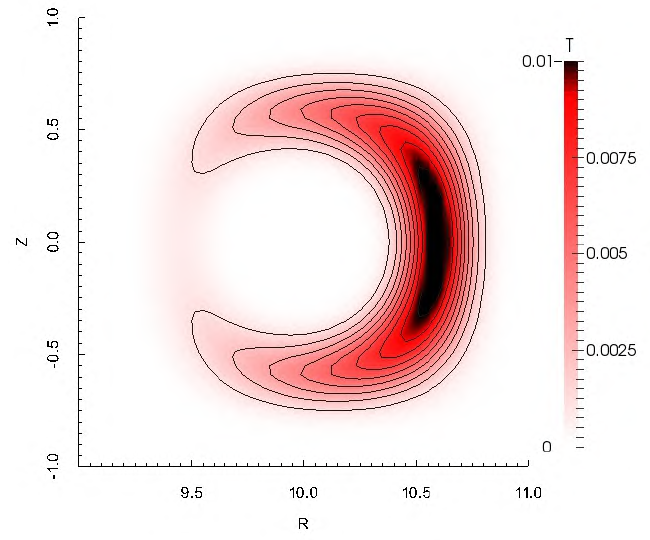}
    \includegraphics[trim = 0px 0px 0px 0px, clip, scale=0.8]{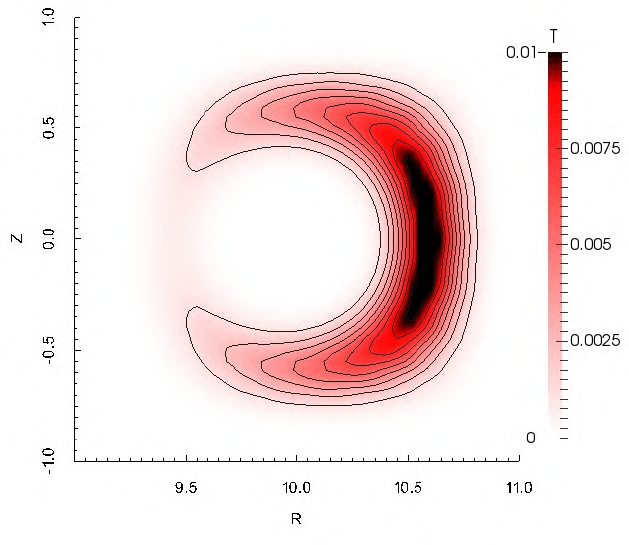}
    \includegraphics[trim = 0px 0px 0px 0px, clip, scale=0.8]{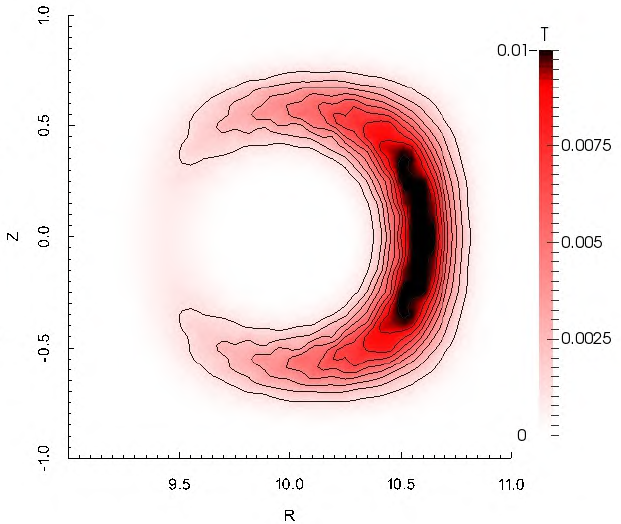}
\caption{These contour plots show the evolution of a three dimensional Gaussian blob at \(t=1\) and \(\varphi = \pi\) for the non-aligned LDG scheme (left), the direct aligned scheme (center) and the self-adjoint aligned scheme (right).
 The latter shows up spurious oscillatory temperature fields. The bean like temperature fields of the aligned discretizations are a consequence of the finite resolution in the direction of \(\varphi\) unlike the non-aligned discretization,
 which has smooth fields even for moderately resolved cases. The resolution was set to  \(P_{RZ} = 3\), \(N_R=N_Z=20\), \(N_\varphi=21 \).} 
 \label{fig:blobs}
\end{figure}

\begin{figure}[!ht]
\centering
\hspace{0.16\textwidth} 
\textbf{D-3}\hspace{0.31\textwidth}\textbf{S-3}\newline
    \includegraphics[trim = 0px 0px 0px 0px, clip, scale=0.55]{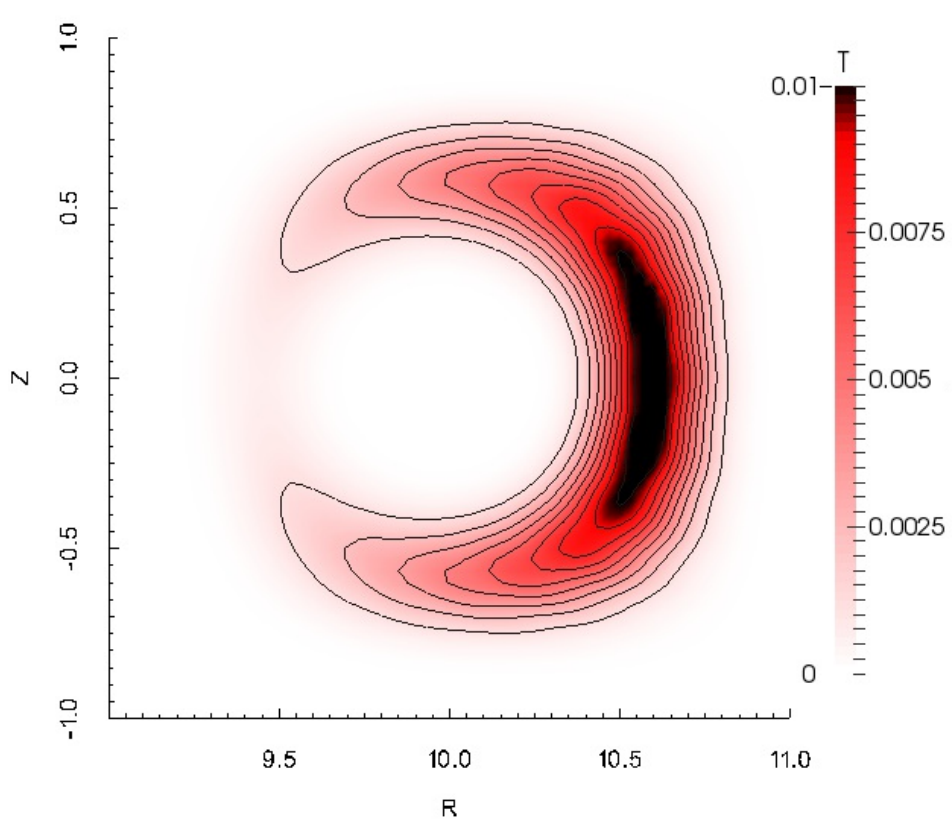}
    \includegraphics[trim = 0px 0px 0px 0px, clip, scale=0.55]{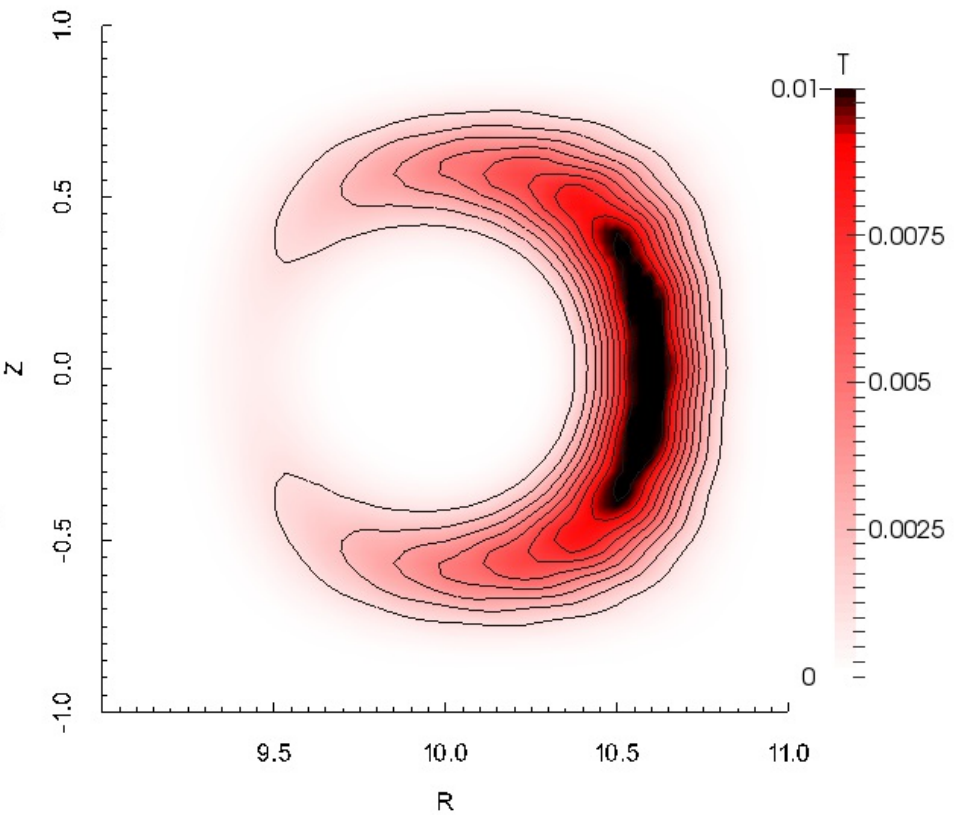}
\caption{Result of simulation of Gaussian blob at $t=1$ and $\varphi=\pi$ for aligned FD schemes. Resolution was set to $N_R=N_Z=60$, $N_\varphi=20$.}
\label{blobsfd}
\end{figure}

\begin{figure}[!ht]
\hspace{-0.13\textwidth} \(a)\)\hspace{0.45\textwidth}\(b)\)\newline
\centering
   \includegraphics[trim = 0px 0px 0px 0px, clip, scale=0.6]{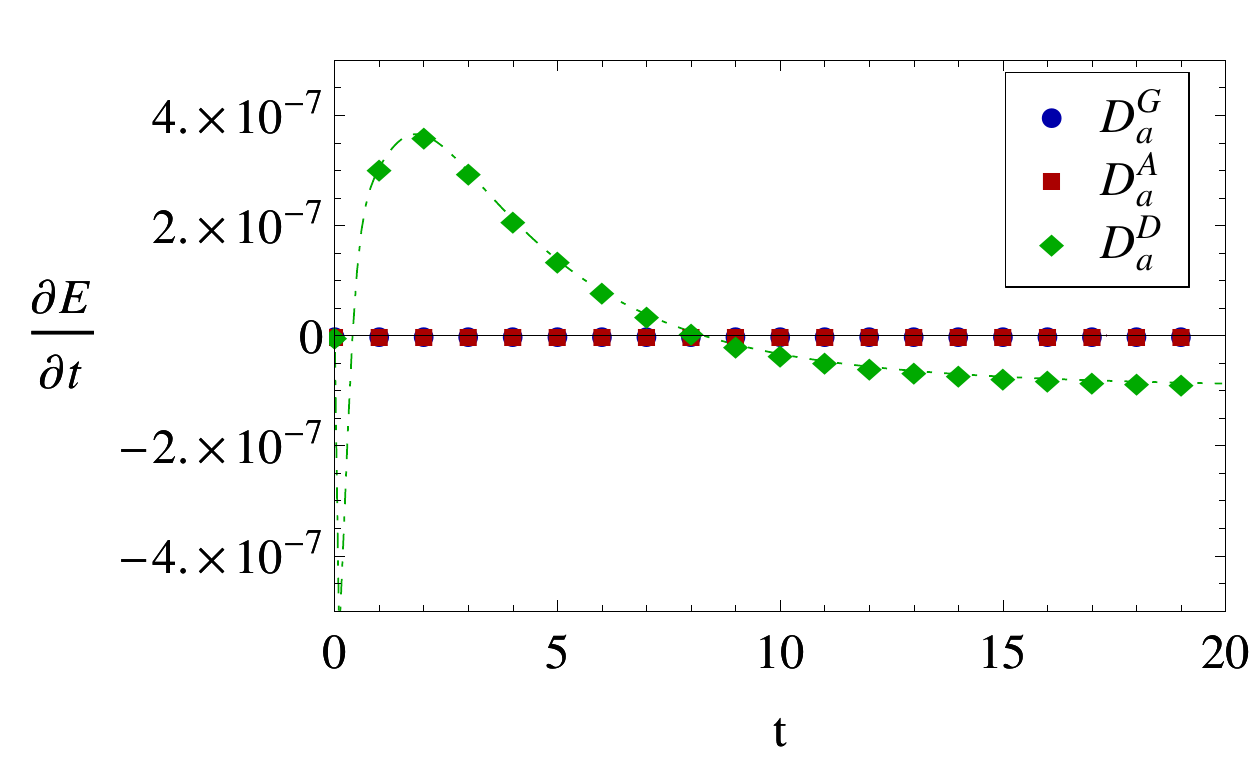}
   \includegraphics[trim = 0px 0px 0px 0px, clip, scale=0.6]{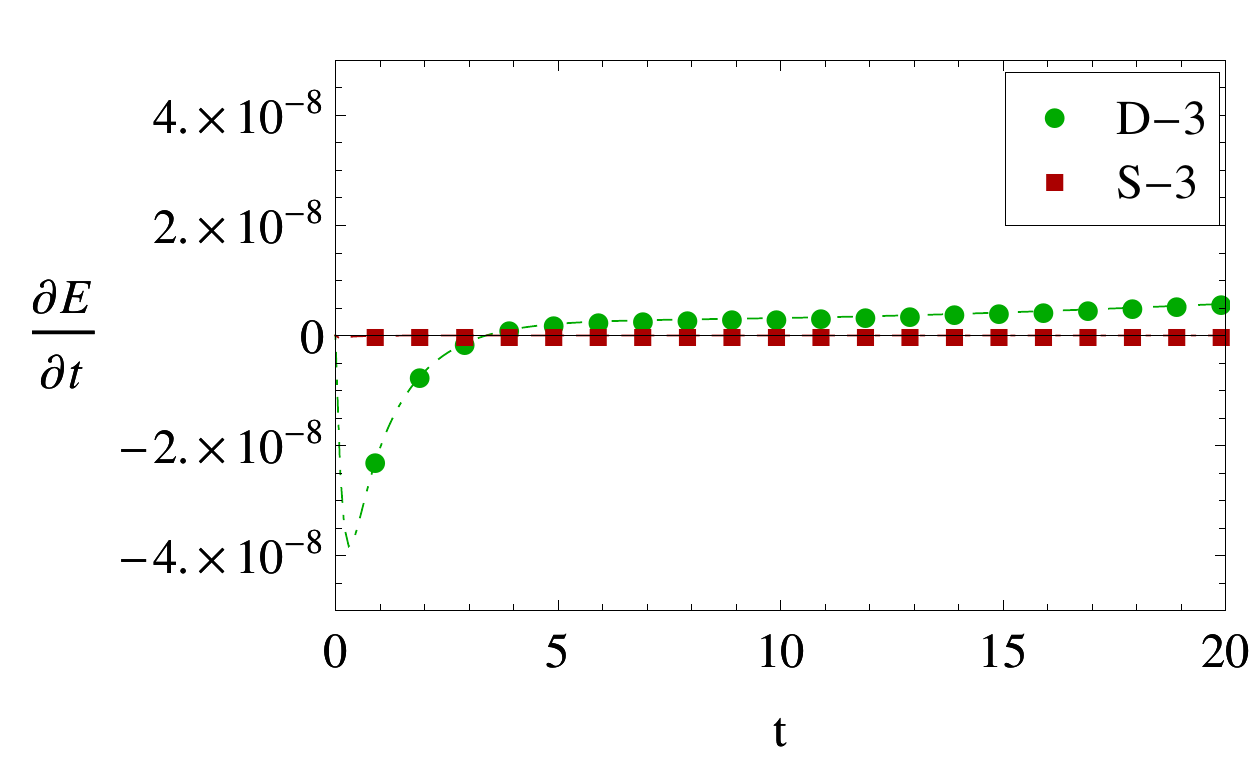}
\caption{a) The energy conservation over time for the averaged non-aligned LDG scheme (blue  circles), the averaged direct aligned scheme (green rhombi) and the averaged self-adjoint aligned scheme (red squares).
 The resolution was set to \(P_{RZ} = 3\), \(N_R=N_Z=20\), \(N_\varphi=20 \).
 b) The energy conservation over time for the \textbf{D-3} scheme (green rhombi) and the \textbf{S-3} scheme (red squares).
 The resolution was set to \(N_R=N_Z=60\), \(N_\varphi=20 \)
 } 
\label{fig:energy}
\end{figure}

We initialize a non-aligned Gaussian blob in \(\varphi\) direction of the form 
\begin{align}\label{eq:gaussian}
 f_0 &= A \exp{\left(-\frac{(R-R_b)^2}{2 \sigma_R^2}-\frac{(Z-Z_b)^2}{2 \sigma_Z^2}-\frac{(\varphi-\varphi_b)^2}{2 \sigma_\varphi^2}\right)}
\end{align}
with $A=0.1$, \(R_b=10.6\), \(Z_b=0.0\), \(\varphi_b = \pi \), \(\sigma_R=\sigma_Z=0.1\), and \(\sigma_\varphi = 0.5\).
To show the convergence of our schemes in time simulations for a general three dimensional perturbation, 
we perform a short highly accurate simulation over \(100\) time steps to \(t=0.01\) with the non-aligned LDG scheme for a resolution of
\(P_{RZ} = 3\), \(N_R=N_Z=51\), \(P_\varphi = 1\)  and \(N_\varphi=162 \). This numerical reference state \(f_{1}\) is used for
the computation of the relative error \( \epsilon_{f_{1}} =  \|f_t- f_{1} \|_{L^2} /\| f_{1} \|_{L^2} \) at \(t=0.01\).
The orders of convergence are listed in Table~\ref{table:blobconv}. The high perpendicular resolution should lead to convergence rates of order two if only the resolution in \(\varphi\) is incremented.
We obtain second order convergence for the non-aligned LDG scheme 
and also for the direct aligned discretization.
However, the self-adjoint aligned discretization has again reduced orders of convergence.
This is contributed to the oscillatory temperature fields, which are depicted in Figure~\ref{fig:blobs}
for a blob with $A=0.1$, \(R_b=10.6\), \(Z_b=0.0\), \(\varphi_b = \pi \), \(\sigma_R=\sigma_Z=0.1\), and \(\sigma_\varphi =2\).
The oscillations arise because the adjoint of the interpolation matrices \(I^{\left\{+,-\right\}}\) do no longer coincide with the three dimensional weights
for all cells. Hence, the transformation is not unitary since \(\vec{\nabla} \cdot \bhat \neq 0\) for the investigated magnetic field. 
For the aligned FD schemes the position of the the interpolation points is always in the center of the sampling points of the interpolating polynomials.
As a consequence the corrugations are better suppressed than for the aligned dG schemes. The initialized structure is not field aligned 
and bean like temperature fields appear for all aligned schemes, whereas the LDG scheme yields smooth results, which is depicted in Figure~\ref{blobsfd}.
 \\
In Figure~\ref{fig:energy}a) and Figure~\ref{fig:energy}b) we show that energy is conserved for all self-adjoint schemes. 
The direct aligned schemes fail in conserving the \(L^1\) norm. We note here that the self-adjoint schemes are also superior in terms of numerical stability. This is based on the fact that the
\(L^2\) norm of the self-adjoint schemes is always \(\langle T, D T \rangle\le 0\) \cite{Stegmeir2015}.
\begin{table}[!ht]
\begin{center}
\caption{\(L^2\) error \(\epsilon_{f_{1}}\) and orders of accuracy 
(\( \textrm{order}_{\Delta \phi}:= \log(L^2 \textrm{error}_{{\Delta \phi}^{-}}/L^2 \textrm{error}_{\Delta \phi})/\log({\Delta \phi}^{-}/{\Delta \phi}) \)) for the averaged non-aligned LDG scheme, the averaged direct aligned scheme and the averaged self-adjoint aligned scheme at \(t=0.01\). The time step is fixed to \(\Delta t = 0.0001\).}
\begin{tabular}{||c|| c c||c c||c c||}
\hline
\multicolumn{1}{||l||}{} & \multicolumn{2}{c||}{ \(D^G_a\) } &  \multicolumn{2}{c||}{\(D^D_a\)} &   \multicolumn{2}{c||}{\(D^A_a\)} \\
\(\Delta \varphi \)&  \(L^2\) error  & order &  \(L^2\) error & order &  \(L^2\) error & order \\ \hline
\(2 \pi/6\) & 2.58E-04 & --    & 7.48E-04 & --   & 7.48E-04 & --  \\ 
\(2 \pi/18\) & 3.50E-05 & 1.82 & 4.62E-04 & 0.44 & 4.64E-04 & 0.44 \\
\(2 \pi/54\) & 3.80E-06 & 2.02 & 5.18E-05 & 1.99 & 1.15E-04 & 1.27 \\
\(2 \pi/162\) & -- & --        & 5.32E-06 & 2.07 & 4.45E-05 & 0.86 \\ \hline
\end{tabular}
\label{table:blobconv}
\end{center}
\end{table}

\subsubsection{Profile maintenance} \label{sec:Profile}
\begin{figure}[!ht]
\hspace{-0.2\textwidth} \(a)\)\hspace{0.45\textwidth}\(b)\)\newline
\centering
   \includegraphics[trim = 0px 0px 0px 0px, clip, scale=0.6]{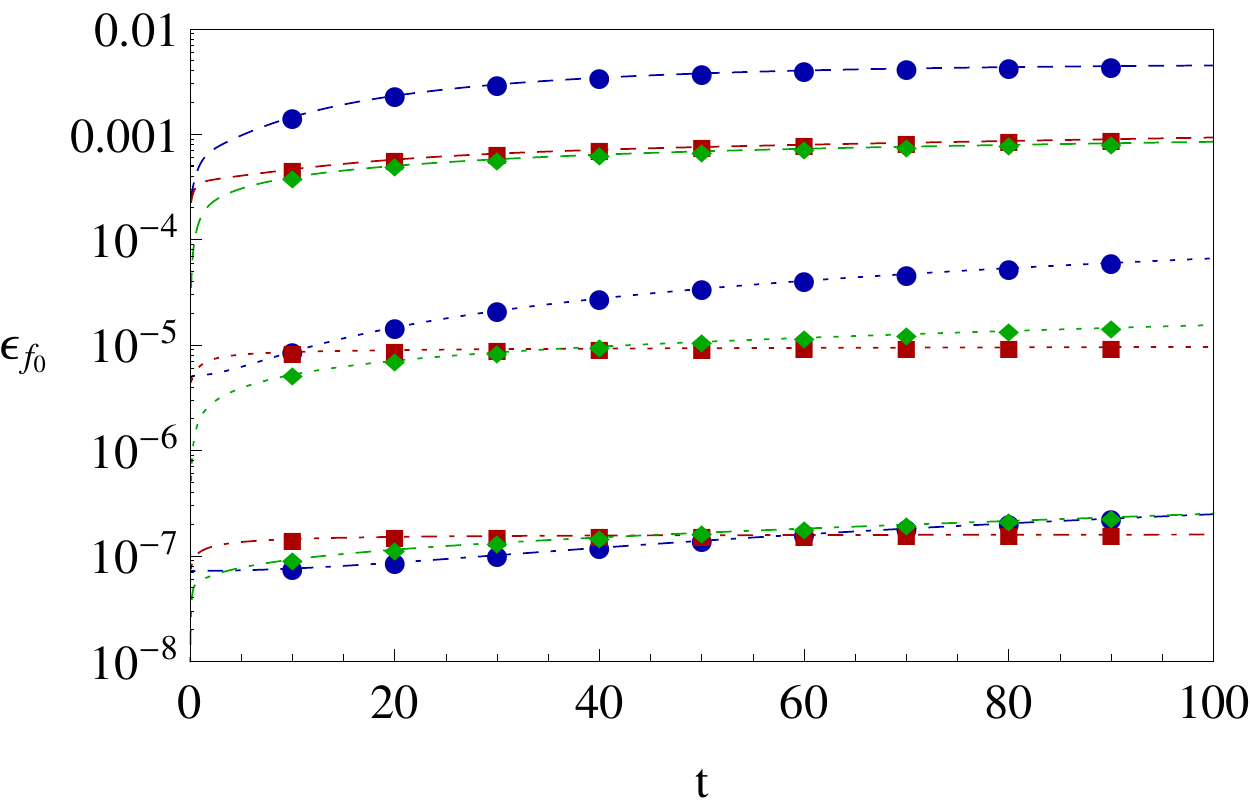}
   \includegraphics[trim = 0px 0px 0px 0px, clip, scale=0.6]{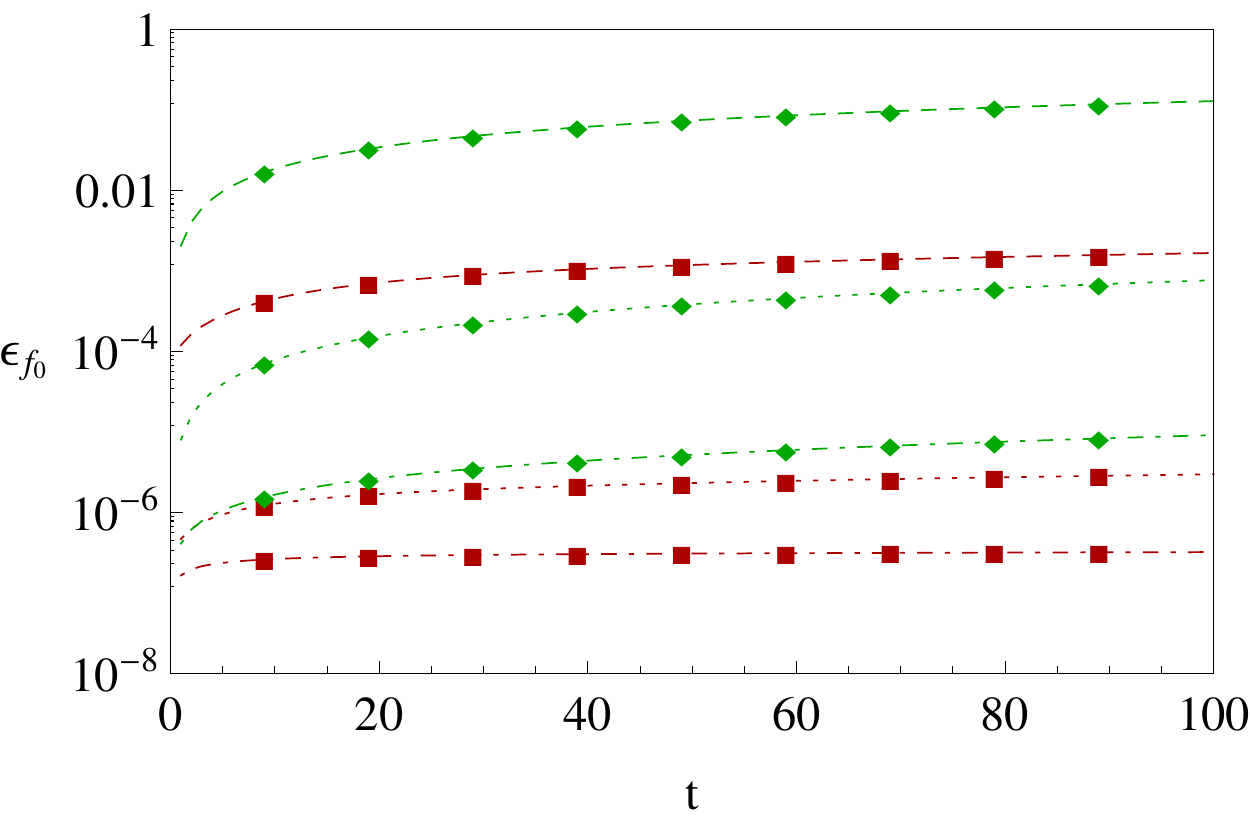}
\caption{a) The relative error \(\epsilon_{f_0}\) over time for the non-aligned LDG scheme (blue  circles), the direct aligned scheme (green rhombi) and the self-adjoint aligned scheme (red squares) 
for three different perpendicular resolutions \(N_R = N_Z=\left\{5,20,80\right\} \) (dashed, dotted, dot-dashed).
b) The relative error \(\epsilon_{f_0}\) over time for the  \textbf{D-3} scheme (green rhombi) and the \textbf{S-3} scheme (red squares) 
for three different perpendicular resolutions \(N_R = N_Z=\left\{15,60,240\right\} \) (dashed, dotted, dot-dashed).
} 
\label{fig:perptimeerror}
\end{figure}

We initialize a profile of the form \( f_0 = 0.1 \psi_p^2 \) for which the parallel Laplacian vanishes exactly to zero \( \Delta_\parallel f_0 = 0\).
Hence, the initial state is conserved over time and deviations are a measure for the numerical 
perpendicular heat flux or numerical sinks and sources. Consequently, we compute the relative error via
\( \epsilon_{f_0} =  \|f_t - f_0 \|_{L^2} / \| f_0 \|_{L^2} \). In Figure~\ref{fig:perptimeerror}a) we analyze its time evolution for three different perpendicular 
resolutions \(N_R = N_Z=\left\{5,20,80\right\} \) fixing 
the resolution in \(\varphi\) to \(N_\varphi=10\) and polynomial coefficients to \(P_\varphi = 1\) and \(P_{RZ} = 3\). 
We observe that the aligned dG schemes are superior to the non-aligned LDG scheme for low resolutions. However, for finer resolutions all three schemes 
do have nearly the same numerical perpendicular heat flux. This error decreases with the order of the interpolating polynomial, which is shown in Table~\ref{table:perperror} at \(t=5\). Here, we observe again superconvergence for the
non-aligned LDG scheme. Furthermore, we note that \( \epsilon_{f_0} \) is independent of \(N_\varphi\) for the non-aligned LDG scheme,
whereas it increases with \(N_\varphi^2\) for the aligned schemes~\cite{Stegmeir2014}. 
This is ascribed to the aligned differences, which are inversely proportional to \( N_\varphi^2 \) for a \(k_\parallel = 0\) mode. \\
For the aligned FD schemes Figure~\ref{fig:perptimeerror}b) shows that the \textbf{D-3} scheme exhibits a much higher
numerical perpendicular heat transport than the \textbf{S-3} scheme, also if \(\vec{\nabla} \cdot \bhat \neq 0 \) (for details cf.~\cite{Stegmeir2015}).
\begin{table}[!ht]
\begin{center}
\caption{\(L^2\) error \(\epsilon_{f_0}\) and orders of accuracy (\( \textrm{order}_{\Delta R}:= \log(L^2 \textrm{error}_{{\Delta R}^{-}}/L^2 \textrm{error}_{\Delta R})/\log({\Delta R}^{-}/{\Delta R})
\)) for the averaged non-aligned LDG scheme, the averaged direct aligned scheme and the averaged self-adjoint aligned scheme}
\begin{tabular}{||c|| c c||c c||c c||}
\hline
\multicolumn{1}{||l||}{} & \multicolumn{2}{c||}{ \(D^G_a\) } &  \multicolumn{2}{c||}{\(D^D_a\)} &   \multicolumn{2}{c||}{\(D^A_a\)} \\
\(\Delta R = \Delta Z\) & \(L^2\) error & order & \(L^2\) error& order & \(L^2\) error & order \\ \hline
2/5 & 9.74E-04 & -- & 3.05E-04 & -- & 4.04E-04 &  -- \\ 
2/10 & 7.22E-05 & 3.75 & 3.11E-05 & 3.29 & 6.22E-05 & 2.70 \\ 
2/20 & 6.22E-06 & 3.54 & 3.91E-06 & 2.99 & 8.13E-06 & 2.94 \\ 
2/40 & 6.44E-07 & 3.27 & 4.89E-07 & 3.00 & 1.01E-06 & 3.00 \\ 
2/80 & 7.31E-08 & 3.14 & 7.75E-08 & 2.66 & 1.36E-07 & 2.90 \\ \hline
\end{tabular} 
\label{table:perperror}
\end{center}
\end{table}
\subsubsection{Aligned blob in tokamak geometry}\label{sec:tokamak}
\begin{figure}[!ht]
\centering
\hspace{0.04\textwidth}\(D^{G,D,A}_a\) (\(t=0\)) \hspace{0.11\textwidth}\(D^G_a\) (\(t=1000\)) \hspace{0.13\textwidth}\(D^D_a\)(\(t=1000\))\hspace{0.13\textwidth}\(D^A_a\)(\(t=1000\))\newline
\includegraphics[trim = 0px 0px 0px 0px, clip, scale=0.26]{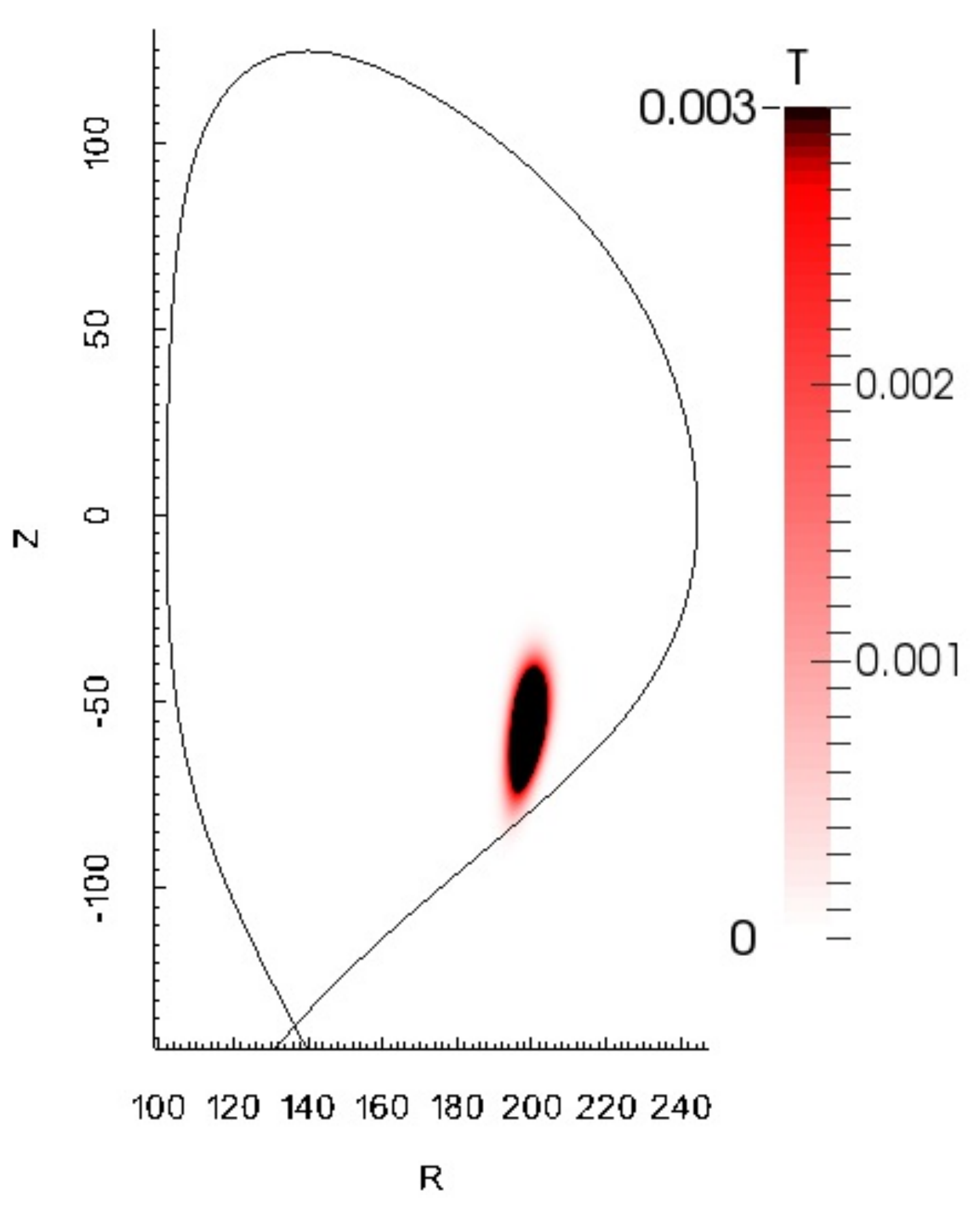}
    \includegraphics[trim = 0px 0px 0px 0px, clip, scale=0.70]{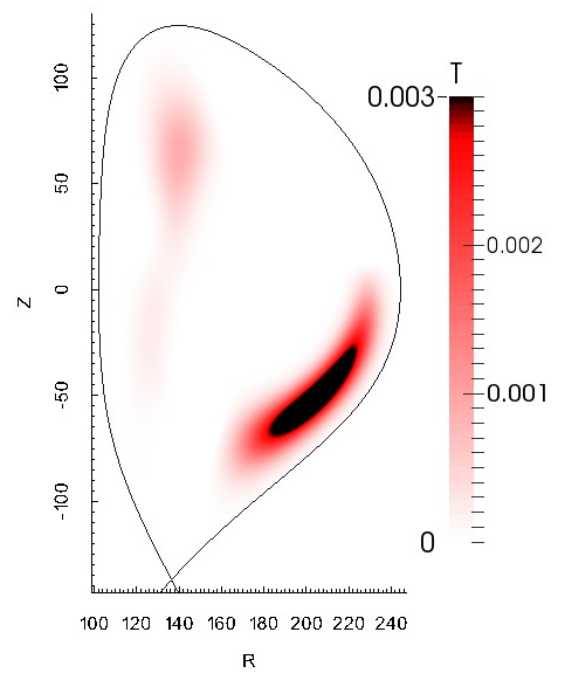}
    \includegraphics[trim = 0px 0px 0px 0px, clip, scale=0.70]{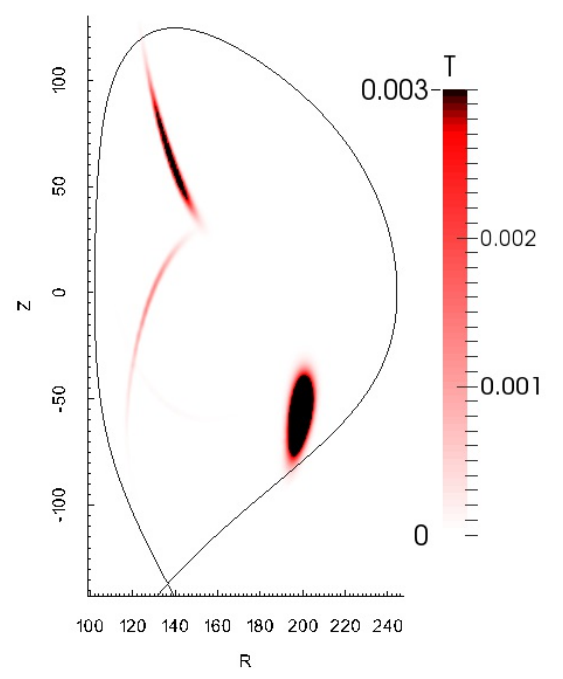}
    \includegraphics[trim = 0px 0px 0px 0px, clip, scale=0.70]{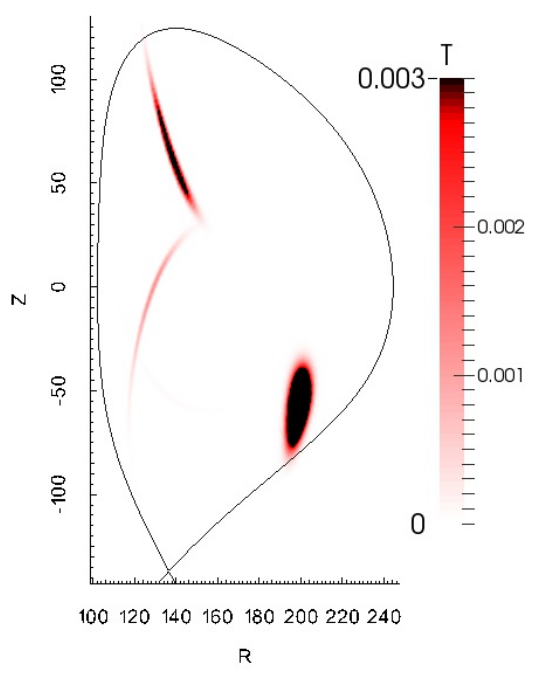}
\caption{
These plots show the evolution of a field aligned blob from \(t=0\) to \(t=1000\) for the non-aligned LDG scheme (2nd column),
the direct aligned scheme (3rd column) and the self-adjoint aligned scheme (4th column). Here, we show the toroidal cross section at \(\varphi = 3\pi/2\).
The non-aligned LDG discretization suffers from pollution by numerical heat fluxes, whereas the aligned dG discretizations are able to retain the initial blob structure whereby also
highly sheared variants emerge in the left top corner of the numerical domain.
The resolution was set to  \(P_{RZ} = 3\), \(N_R=80\), \(N_Z=120\), \(P_\varphi = 1\)  and \(N_\varphi=20 \)..
 } 
 \label{fig:alignedblobs}
 \end{figure}

\begin{figure}[!ht]
\centering
    \includegraphics[trim = 0px 0px 0px 0px, clip, scale=0.6]{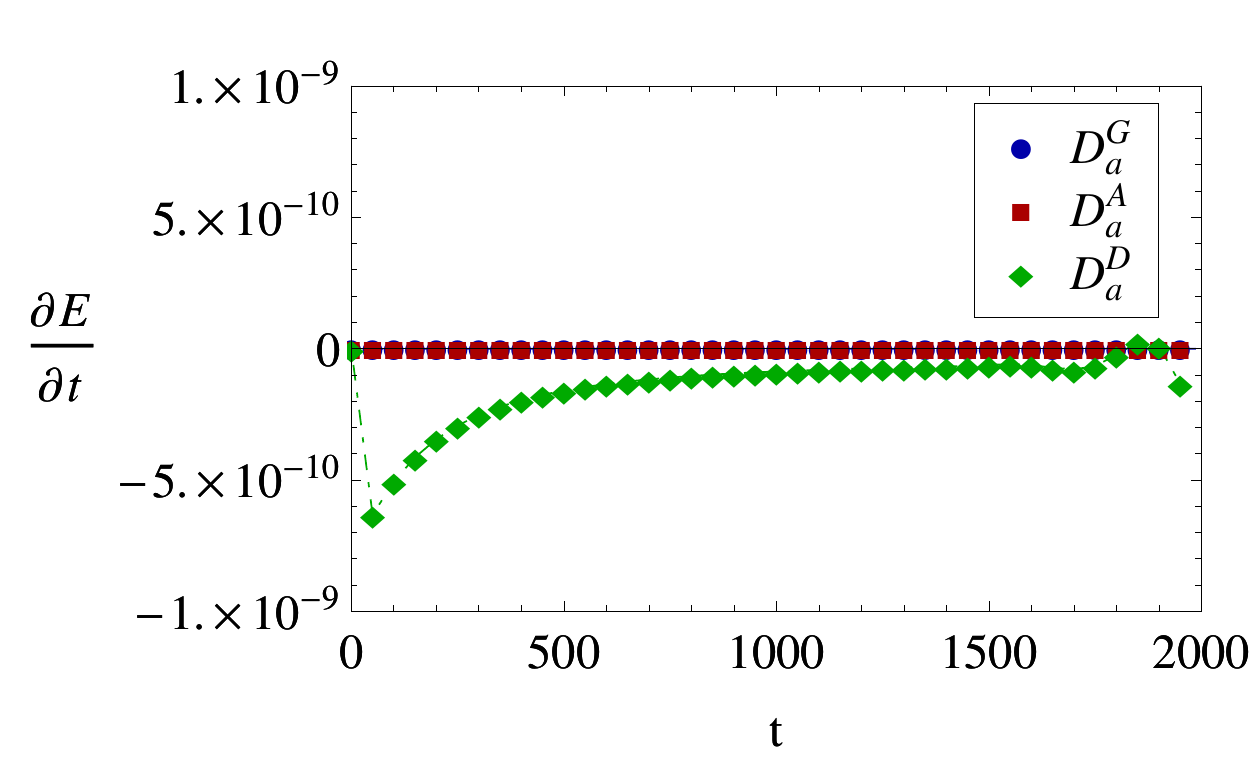}    
\caption{
The energy conservation over time of a field aligned Gaussian blob 
for the averaged non-aligned LDG scheme (blue circles), the averaged direct aligned scheme (green rhombi) and the averaged self-adjoint aligned scheme (red squares) is shown.
 } 
   \label{fig:xdedt}
\end{figure}

We verify our schemes now on a realistic axisymmetric tokamak equilibrium with lower X-point. The used Solov'ev equilibrium is an analytical solution of the Grad-Shafranov equation in
cylindrical coordinates \cite{Cerfon2010} and resembles to a sufficient extent a typical deuterium discharge in the COMPASS tokamak with major radius \(R_0  = 0.56 m\), inverse aspect ratio 
\(\epsilon = 0.41\), elongation \(\kappa = 1.75\), triangularity \(\delta=0.47\), toroidal magnetic field \(B_0 = 1T\) and reference electron Temperature \(T_{e0} = 500eV\)\cite{wesson2011}.
This fixes the drift scale \(\rho_{s0}=\sqrt{m_i T_e}/e B_0\) and the ion gyro-frequency \(\Omega_{i}= e B_0/m_i\), which normalize the spatial and time scales.
The corresponding geometric coefficients are 
\begin{align}
A&=0, &
c_1 &= 0.0735011444550040364542354505071,\nonumber \\
c_2 &= -0.0866241743631724894540195833599,&
c_3 &= -0.146393154340110145566533995171, \nonumber\\
c_4 &= -0.0763123710053627236722106917166, &
c_5 &=0.0903179011379421292953276540381,\nonumber\\
c_6 &= -0.0915754123901870980044268057017, &
c_7 &= -0.00389228297983755811356246513352, \nonumber\\
c_8 &=0.0427189122507639110238025727235, &
c_9 &=0.227554564600278024358552580518, \nonumber\\
c_{10}&=-0.130472413601776209642913874538,& 
c_{11}&= -0.0300697410847693691591393359449, \nonumber\\
c_{12} &= 0.00421267189210391228482454852824.
\end{align}
In Figure~\ref{fig:alignedblobs} we show the time evolution of an initially field aligned blob with \(k_\parallel\neq0 \) at \(\varphi = 3 \pi /2\) for a typical spatial resolution of 
\(P_{RZ} = 3\), \(N_R=80\), \(N_Z=120\), \(P_\varphi = 1\)  and \(N_\varphi=20 \). 
 We initialize a two dimensional Gaussian blob at \(\varphi= \pi\) of the form \(f_{0}(R,Z,\pi)\) (see Eq.~\eqref{eq:gaussian})
and with parameters 
\(A=0.1\), 
\(R_b=R_0  \left(1 + 0.8 \epsilon \right) \), 
\(Z_b= 0.0  \), 
\(\varphi_b= \pi\), 
\(\sigma_R=\sigma_Z = 5 \rho_{s0} \). With the help of the field line transformation

\begin{align}
 f_{0,\parallel}(R,Z,\pi \pm\Delta\varphi) & =\exp{\left(-\frac{\Delta\varphi^2}{2 \sigma_\varphi^2}\right)} T^{\mp1}_{\Delta\varphi} f_0(R,Z,\pi) 
\end{align}
we transform the two dimensional intial field to all other planes. With \(\sigma_\varphi=0.5 \pi R_0\) this yields a field aligned blob that is spread over approximately half of the toroidal domain. 
Due to the sheared magnetic field the two dimensional Gaussian blob structure is distorted at \(\varphi \neq \pi\).
As time progresses we expect that the initial cross section of the field aligned blob is conserved. Due to parallel heat conduction the field aligned blob traces out the magnetic field lines and 
highly sheared variants can emerge on the same toroidal plane. 
As depicted in Figure~\ref{fig:alignedblobs} this is not the case for the non-aligned LDG scheme. 
Here, the coarse resolution fails to resolve the complex flute-mode properly as time advances. The \(k_\parallel\neq 0 \) flute-mode is dominated by numerical heat fluxes. 
Only the aligned dG schemes can resolve this mode at reasonably low resolution in \(\varphi\) direction.
The high numerical heat flux of the non-aligned LDG scheme is caused by the fact that field lines are sampled by the grid. 
In contrast aligned schemes resolve the magnetic field lines with a high accuracy based on a highly accurate field line tracing procedure.
We remark again the violation of the \(L^1\) norm conservation of the direct aligned dG scheme in contrast to the self-adjoint schemes. For this choice of numerical parameters
this violation becomes significant at \(t\approx 1800\), which is shown in Figure~\ref{fig:xdedt}.

\section{Conclusion} \label{sec:conclusion}
We have presented three different dG discretizations for the parallel diffusion operator. 
The non-aligned LDG discretization is flexible in the order of convergence whilst keeping the numerical
perpendicular heat fluxes at a small level for a basic \(k_\parallel = 0\) mode and conserving energy exactly.
Our numerical measurements revealed that in general \(\vec{\nabla} \cdot \bhat \neq 0\) cases the aligned dG schemes may exhibit 
stagnating, reduced or irregular convergence rates whereas the non-aligned LDG scheme remains robust.
However, especially for flute-modes and coarse resolutions the aligned schemes are advantageous to the non-aligned LDG scheme.
Most notably the non-aligned LDG discretization requires significantly higher resolutions for \(k_\parallel \neq 0\) flute-modes in typical X-point geometry.
We compared the latter dG schemes to the aligned FD schemes of References~\cite{Stegmeir2014,Stegmeir2015} for a magnetic field with \(\vec{\nabla} \cdot \bhat \neq 0\). 
The direct aligned FD scheme converges with robust rates but is inferior to the self-adjoint aligned FD scheme regarding energy conservation and flute mode maintenance. 
On the downside the latter scheme showed up degraded and anomalous convergence rates due to the appearance of few corrugations on the grid scale.
The herein presented dG discretizations for the parallel diffusion operator are implemented into a \textbf{F}ull-F \textbf{EL}ectromagnetic model in \textbf{TOR}oidal geometry (\textbf{FELTOR}) and
support the computation of long term turbulence simulations of X-point equilibria. Those results are postponed to future publications.

\section*{Acknowledgements}
The authors want to thank Alexander Kendl and Lukas Einkemmer (U Innsbruck) and Farah Hariri (EPFL Lausanne) for contributing to this work with fruitful comments.
This work was supported by the Austrian Science Fund (FWF) Y398. Matthias Wiesenberger acknowledges further support by a \"OAW KKK\"O Impulsprojekt. 
This work has been carried out within the framework of the EUROfusion Consortium and has received funding
from the Euratom research and training programme 2014-2018 under grant agreement No 633053. The views and opinions expressed herein do not necessarily reflect those of the European Commission. 
The computational results presented have been achieved in part using the Vienna Scientific Cluster (VSC). 
The research leading to these results has partly received funding from the European Research Council under the European Union’s Seventh Framework Programme (FP7/2007-2013)/ERC Grant Agreement No. 277870. 
A part of this work was carried out using the HELIOS supercomputer system at Computational Simulation Centre of International Fusion Energy Research Centre (IFERC-CSC), 
Aomori, Japan, under the Broader Approach collaboration between Euratom and Japan, implemented by Fusion for Energy and JAEA.

\bibliography{refs}


\end{document}